\pdfoutput=1
\documentclass[fleqn]{aa}

\usepackage{amsmath,amssymb}
\usepackage{txfonts}
\usepackage{natbib}
\bibpunct{(}{)}{;}{a}{}{,} 
\usepackage[colorlinks=True,
            urlcolor=blue,
            linkcolor=blue,
            citecolor= blue,
	    pdftex]{hyperref}
\usepackage{booktabs}
\usepackage{longtable}
\usepackage[pdftex]{graphicx}

\newcommand{\lp}{\left(}
\newcommand{\rp}{\right)}
\newcommand{\tb}{\bar{T}}
\newcommand{\rb}{\bar{\rho}}

\begin{document}

\title{Dipolar versus multipolar dynamos: the influence of the background
density stratification}

\author{T.\ Gastine \and L.\ Duarte \and J.\ Wicht}
\institute{Max Planck Institut f\"ur Sonnensystemforschung, Max Planck
Stra{\ss}e 2, 37191 Katlenburg Lindau, Germany \email{gastine@mps.mpg.de}}


\date{\today}

\abstract
{Dynamo action in giant planets and rapidly rotating stars leads to a
broad variety of magnetic field geometries including
small scale multipolar and large scale dipole-dominated topologies.
Previous dynamo models suggest that solutions become multipolar
once inertia becomes influential. Being tailored for terrestrial
planets, most of these models neglected the background
density stratification.}
{We investigate the influence of the density stratification on
convection-driven dynamo models.}
{Three-dimensional nonlinear simulations of rapidly rotating
spherical shells are employed using the anelastic approximation to
incorporate density stratification. A systematic parametric
study for various density stratifications and Rayleigh numbers
at two different aspect ratios allows to explore the
dependence of the magnetic field topology on these parameters.}
{Anelastic dynamo models tend to produce a broad range of magnetic field
geometries that fall on two distinct branches with either strong
dipole-dominated or weak multipolar fields.
As long as inertia is weak, both branches can coexist but the dipolar branch
vanishes once inertia becomes influential.
The dipolar branch also vanishes for stronger density stratifications. The
reason is the concentration of the convective columns in a narrow
region close to the outer boundary equator, a configuration that
favors non-axisymmetric solutions.
In multipolar solutions, zonal flows can become significant and participate in
the toroidal field generation. Parker dynamo waves may then play
an important role close to onset of dynamo action leading to a
cyclic magnetic field behavior.}
{These results are compatible with the magnetic field of gas planets that
are likely generated in their deeper conducting envelopes where the density
stratification is only mild.
Our simulations also suggest that the fact that late M dwarfs have dipolar or
multipolar magnetic fields can be explained in two ways. They may differ either
by the relative influence of inertia or fall into the regime where both types of
solutions coexist.
}


\maketitle

\section{Introduction}

The magnetic fields of planets and rapidly rotating low-mass stars are
maintained by
magnetohydrodynamic dynamos operating in their interiors. Scaling laws for
convection-driven dynamos successfully predict the magnetic field strengths
for both types of objects indicating that similar
mechanisms are at work \citep{Christensen09}. Recent observations show a broad
variety of magnetic
field geometries, ranging from large-scale dipole-dominated topologies \citep[as
on Earth, Jupiter and some rapidly rotating M dwarfs, see
e.g.][]{Donati06,midM} to
small-scale more complex magnetic structures \citep[e.g.][]{earlyM,lateM}.
Explaining the different field geometries remains a prime goal of
dynamo theory. Unfortunately, the extreme parameters
of planetary and stellar dynamo regions can not directly be adopted
in the numerical models so that scaling laws are of prime importance
\citep{Christensen10}.

Global simulations of rapidly rotating convection successfully reproduce many
properties of planetary dynamos \citep[e.g.][]{Christensen06}. They show that 
the
ordering influence of the Coriolis force is responsible for producing
a dominant large scale dipolar magnetic field \citep[for a stellar application, see
also][]{Brown10} provided the Rayleigh number is not too large.
Multipolar geometries are assumed once the Rayleigh number
is increased beyond a value where inertial forces become important
\citep{Kutzner02,Sreenivasan06}.
\cite{Christensen06} suggest that the ratio of (nonlinear) inertial to
Coriolis forces can be quantified with what they call the ``local Rossby
number''
$\text{Ro}_\ell=u_{\text{rms}}/\Omega \ell$,
where $\ell$ is the typical lengthscale of the convective flow.
Independently of the other system parameters,
the transition between dipole-dominated and multipolar magnetic fields always
happens around $\text{Ro}_\ell \simeq 0.1$.
Scaling laws then, for example, predict a dipole-dominated field for Jupiter
and a multipolar field for Mercury \citep{Olson06}.
The latter may explain the strong quadrupolar component in the planet's
magnetic field \citep{Christensen06nat}.
Earth lies at the boundary where the field is dipolar most of the time
but occasionally forays into the multipolar regime allowing for magnetic field
reversals.

The scaling laws based on Boussinesq simulations are geared to model the dynamo
in Earth's liquid metal core where the background density and temperature
variations can be neglected. Their application to gas giants and
stars where both quantities vary by orders of magnitude is therefore
questionable \citep{Chabrier06,Nettelmann12}.
Recent compressible dynamo models of fully-convective stars
suggest a strong influence of the density stratification on the geometry of the
magnetic field: while the fully compressible models of \cite{Dobler06} have a
significant dipolar component, the strongly stratified anelastic models of
\cite{Browning08} tend to produce multipolar dynamos \citep[see
also][]{Bessolaz11}. These differences stress the need of more systematic
parameter studies to clarify the influence of density stratification.

In addition, most of the previous Boussinesq studies have employed rigid flow
boundary conditions for modelling terrestrial dynamo models.
Stress-free boundary conditions, more appropriate for
gas planets and stars, show a richer dynamical
behaviour, including hemispherical dynamos and bistability
where dipole-dominated and multipolar
dynamos coexist at the same parameters
\citep[e.g.][]{Grote00,Busse06,Goudard08,Simitev09,Sasaki11}. This
challenges the usefulness of the $\text{Ro}_\ell$ criterion since
bistable cases exist significantly below $\text{Ro}_\ell\simeq 0.1$
\citep[e.g.][]{Schrinner12}.

Here we adopt the anelastic approximation \citep[e.g.][]{Glatz1,Brag95,Lantz99}
to explore the effect of a background density stratification on the
dynamo process while filtering out fast acoustic waves. We conduct
an extensive parameter study where we vary the degree of stratification and
the Rayleigh number to determine the parameter range where dipole-dominated
fields can be expected. The anelastic approximation and the numerical methods
are introduced in section~\ref{sec:model}. In section~\ref{sec:results}, we
present the results of the parametric study and describe the different dynamo
regimes before relating our results to observations in the concluding
section~\ref{sec:discussion}.

\section{The dynamo model}
\label{sec:model}

\subsection{An anelastic formulation}

We consider MHD simulations of a conducting ideal gas in a spherical shell
rotating at a constant frequency $\Omega$ about the $z$ axis. Convective
motions are driven by a fixed entropy contrast $\Delta s$ between
the inner radius $r_i$ and the outer radius $r_o$. Following the previous
parametric studies of \cite{Christensen06}, we use a
dimensionless formulation, where the shell thickness $d=r_o-r_i$ is the
reference lengthscale, the viscous diffusion time $d^2/\nu$ is the
reference timescale and  $\sqrt{\rho_{\text{top}} \mu \lambda \Omega}$ is
the magnetic scale. Density and temperature are non-dimensionalised using
their values at the outer boundary  $\rho_{\text{top}}$ and
$T_{\text{top}}$ and $\Delta s$ serves as the entropy scale.
The kinematic viscosity $\nu$, magnetic diffusivity $\lambda$,
magnetic permeability $\mu$, thermal diffusivity $\kappa$ and
heat capacity $c_p$ are assumed to be constant.

Following the anelastic formulation by
\cite{Glatz1}, \cite{Brag95} and \cite{Lantz99} we adopt a non-magnetic,
hydrostatic and adiabatic background reference state that we
denote with overbars in the following.
It is defined by the temperature profile $d\tb/dr
= -g/c_p$ and the assumption of a polytropic gas $\rb(r) = \tb^m$,
where $m$ is the polytropic index.
Gravity typically increases linearly with radius in
Boussinesq models where the density is approximately constant.
Many anelastic models on the other hand assume that the density is predominantly
concentrated below the dynamo zone so that $g\propto 1/r^2$ provides
a better approximation \citep[e.g.][]{Glatz84,Jones11}.
We adopt the form

\begin{equation}
 g(r) = g_1 \dfrac{r}{r_o} + g_2 \dfrac{r_o^2}{r^2},
\end{equation}
with constants $g_1$ and $g_2$ to allow for both alternatives which then
leads to the following background temperature profile
\citep[see also][]{Jones11,Gastine12}

\begin{equation}
 \tb(r) = c_o\lp g_2\dfrac{r_o}{r}-\dfrac{g_1}{2} \dfrac{r^2}{r_o^2}\rp+
1-c_o \lp g_2 -\dfrac{g_1}{2} \rp,
\end{equation}
with

\begin{equation}
 c_o = \dfrac{\eta \lp \exp\frac{N_\rho}{m}-1 \rp
}{\frac{g_1}{2}(1-\eta^2)\eta+g_2(1-\eta)}
 \text{~and~} N_\rho = \ln\dfrac{\rb(r_i)}{\rb(r_o)}.
\end{equation}
Here, $\eta=r_i/r_o$ is the aspect ratio of the
spherical shell and $N_\rho$ is the number of density scale
heights covered by the density background.

Under the anelastic approximation the dimensionless equations governing
convective motions and magnetic field generation are then given by

\begin{equation}
\begin{aligned}
   \text{E}\lp\dfrac{\partial \vec{u}}{\partial
   t}+\vec{u}\cdot\vec{\nabla}\vec{u}\rp
    = &
 -\vec{\nabla}{\dfrac{p}{\rb}}+\dfrac{\text{Ra}\,\text{E}}{\text{Pr}
}g(r) s\,\vec{e_r}
-2\vec{e_z}\times\vec{u}
\\  & + \dfrac{1}{\text{Pm}\ \rb}\lp\vec{\nabla}\times
\vec{B}\rp \times \vec{B}
+  \dfrac{\text{E}}{\rb} \vec{\nabla}\cdot\tens{S},
 \end{aligned}
 \label{eq:NS}
\end{equation}
\begin{equation}
 \dfrac{\partial \vec{B}}{\partial t} = \vec{\nabla}\times \lp
\vec{u}\times\vec{B}\rp
 +\dfrac{1}{\text{Pm}}\vec{\Delta}\vec{B},
 \label{eq:induction}
\end{equation}
\begin{equation}
\begin{aligned}
\rb\tb\lp\dfrac{\partial s}{\partial t} + \vec{u}\cdot\vec{\nabla} s\rp = &
\dfrac{1}{\text{Pr}}\vec{\nabla}\cdot\lp\rb\tb \vec{\nabla} s\rp +
\dfrac{\text{Pr}}{\text{Ra}}\,\dfrac{c_o}{2\rb} (1-\eta)\,\tens{S}^2 \\
& + \dfrac{\text{Pr}}{\text{Ra\,E\,Pm}^2}\, c_o (1-\eta)
\lp\vec{\nabla}\times\vec{B}\rp^2,
\end{aligned}
\label{eq:entropy}
\end{equation}
\begin{equation}
 \vec{\nabla}\cdot \lp \rb\vec{u} \rp = 0  \quad,\quad  \vec{\nabla}\cdot
\vec{B}  = 0
 \label{eq:anel},
\end{equation}
where $\vec{u}$, $\vec{B}$, $p$  and $s$ are velocity, magnetic field, pressure
and entropy disturbance of the background state, respectively.
$\tens{S}$ is the traceless rate-of-strain tensor
with a constant kinematic viscosity given by

\begin{equation}
\tens{S}=\rb\left(\frac{\partial u_i}{\partial x_j} +
\frac{\partial u_j}{\partial x_i}-\frac{2}{3} \delta_{ij}
\vec{\nabla}\cdot\vec{u}\right),
\label{eq:tenseur}
\end{equation}
where $\delta_{ij}$ is the identity matrix.
In addition to the aspect ratio of the spherical shell $\eta$ and the two
parameters involved in the description of the reference state ($N_\rho$ and $m$),
the system of equations (\ref{eq:NS}-\ref{eq:anel}) is governed by
four dimensionless parameters, namely the Ekman number E, the Prandtl number Pr,
the magnetic Prandtl number Pm and the Rayleigh number Ra defined by

\begin{equation}
 \text{E} = \dfrac{\nu}{\Omega d^2} \ ;\ \text{Pr} =
\dfrac{\nu}{\kappa} \ ;\  \text{Pm} = \dfrac{\nu}{\lambda}
\ ;\ \text{Ra} = \dfrac{g_{\text{top}} d^3 \Delta s}{c_p \nu\kappa},
\label{eq:params}
\end{equation}
where $g_{\text{top}}$ is the reference gravity at the outer boundary $r_o$.

\subsection{The numerical method}

The numerical simulations in this parameter study have been computed with
the anelastic version of the code MagIC \citep{Wicht02,Gastine12}, which has
been validated in an anelastic dynamo benchmark \citep{Jones11}. To
solve the system of equations (\ref{eq:NS}-\ref{eq:anel}),
$\rb\vec{u}$ and $\vec{B}$ are decomposed into poloidal and toroidal
contributions

\begin{equation}
 \rb \vec{u} = \vec{\nabla}\times(\vec{\nabla}\times W \vec{e_r}) +
\vec{\nabla}\times Z \vec{e_r} \ ,\ \vec{B}=
\vec{\nabla}\times(\vec{\nabla}\times C \vec{e_r}) +
\vec{\nabla}\times A \vec{e_r}.
\end{equation}
The unknowns $W$, $Z$, $A$, $C$, $s$ and $p$ are then expanded in spherical harmonic
functions up to degree $\ell_{\text{max}}$ in colatitude $\theta$
and longitude $\phi$ and in Chebyshev polynomials up to degree $N_r$ in the radial
direction. An exhaustive description of the complete numerical method and the
associated spectral transforms can be found in \citep{Glatz1}.
Typical numerical resolutions employed in this study range from
($N_r=61,\,\ell_{\text{max}}=85$) for Boussinesq simulations close to onset to
($N_r=81,\,\ell_{\text{max}}=133$) for the more demanding cases with a
significant density contrast.

In all the simulations presented in this study, we have assumed
constant entropy boundary conditions at $r_i$ and $r_o$.
The mechanical boundary conditions are either no slip at the inner and 
stress-free at the outer boundary or stress-free at both limits (see
Tab.~\ref{tab:results}). The magnetic field is
match to a potential field at both boundaries.

\subsection{Non-dimensional diagnostic parameters}

To quantify the impact of the different control parameters on magnetic
field and flow, we analyse several diagnostic non-dimensional properties.
The typical rms flow amplitude in the shell is either given as the magnetic
Reynolds number $\text{Rm} = u_{\text{rms}} d/\lambda$ or the Rossby number
$\text{Ro}=u_{\text{rms}}/\Omega d$. Rm is a measure for the ratio of
magnetic field generation to Ohmic dissipation and thus an important
quantity for any dynamo. The Rossby number quantifies the ratio between inertia
and Coriolis forces but
\citet{Christensen06} demonstrated
that the local Rossby number $\text{Ro}_\ell=u_{\text{rms}}/\Omega \ell$
is a more appropriate measure, at least concerning the impact of
inertia on the magnetic field geometry. The typical flow lengthscale $\ell$
can be calculated based on the spherical harmonic flow contributions

\begin{equation}
\ell = \pi/\bar{\ell}_u,
\end{equation}
with

\begin{equation}
 \bar{\ell}_u = \sum_\ell   \dfrac{\ell\ \langle \vec{u}_\ell\cdot\vec{u}_\ell
\rangle}{\langle  \vec{u}\cdot \vec{u} \rangle}.
\end{equation}
Here, $\vec{u}_\ell$ is the velocity field at a given spherical harmonic degree $\ell$
and $\langle \cdots \rangle$ corresponds to an average over time and radius.

The magnetic field strength can, for example, be measured by
the Elsasser number $\Lambda = B_{\text{rms}}^2/\rho\mu_o\lambda \Omega$,
which is supposed to provide an estimate for the ratio of Lorentz to
Coriolis forces.
This estimate may be rather far from the true force balance
\citep[e.g.][]{Wicht10}. The modified Elsasser number suggested by
\cite{King12} and defined by

\begin{equation}
 \Lambda_{\ell} = \dfrac{\Lambda}{\text{Rm}} \dfrac{d}{\delta_B}
\quad\text{with}\quad \delta_B = \dfrac{\pi
d}{2\sqrt{\bar{\ell}_B^2+\bar{m}_B^2} },
 \label{eq:elsloc}
\end{equation}
provides a more appropriate measure, being more directly related to the
ratio of the two respective terms in the Navier-Stokes
equation (\ref{eq:NS}) when approximating the lengthscale entering in the curl
in the Lorentz force by $\delta_B^{-1}$. In addition to the
typical spherical harmonic degree $\bar{\ell}_B$, the typical order $\bar{m}_B$
is also used since both seem to contribute \citep{King12}:

\begin{equation}
 \bar{\ell}_B = \sum_\ell   \dfrac{\ell\ \langle \vec{B}_\ell\cdot\vec{B}_\ell
\rangle}{\langle  \vec{B}\cdot \vec{B} \rangle} \ ,\
 \bar{m}_B = \sum_m \dfrac{m\ \langle \vec{B}_m\cdot\vec{B}_m
\rangle}{\langle  \vec{B}\cdot \vec{B} \rangle}.
\end{equation}

Finally, the geometry of the magnetic field is quantified by its dipolarity

\begin{equation}
 f_{\text{dip}} =
\dfrac{\oiint\vec{B}^2_{\ell=1,m=0} \sin\theta\,d\theta\,d\phi}
{\oiint\vec{B}^2\sin \theta\,d\theta\,d\phi},
\end{equation}
that measures the ratio of the magnetic energy of the dipole to
the total magnetic energy at the outer boundary $r_o$.

\subsection{A parameter study}

\begin{table}
\caption{Critical Rayleigh numbers and corresponding azimuthal wave numbers for
the simulations considered in this study (for all of them $\text{E}=10^{-4}$
and $\text{Pr}=1$).}
\centering
\begin{tabular}{cccc}
  \toprule
  Aspect ratio & $N_\rho$ & $\text{Ra}_c$ & $m_{\text{crit}}$ \\
  \midrule
  0.2 & 0   & $8.706\times 10^{5}$ & 4  \\
  0.2 & 0.5 & $1.378\times 10^{6}$ & 5  \\
  0.2 & 1   & $1.935\times 10^{6}$ & 5  \\
  0.2 & 1.5 & $2.617\times 10^{6}$ & 5  \\
  0.2 & 1.7 & $2.929\times 10^{6}$ & 6 \\
  0.2 & 2   & $3.455\times 10^{6}$ & 6  \\
  0.2 & 2.5 & $4.591\times 10^{6}$ & 6  \\
  0.2 & 3   & $4.647\times 10^{6}$ & 43 \\
 \midrule
  0.6 & 0.01& $1.737\times 10^{5}$ & 21 \\
  0.6 & 0.5 & $3.126\times 10^{5}$ & 28 \\
  0.6 & 1   & $5.174\times 10^{5}$ & 34 \\
  0.6 & 1.5 & $8.151\times 10^{5}$ & 39 \\
  0.6 & 2   & $1.140\times 10^{6}$ & 53 \\
  0.6 & 3   & $1.530\times 10^{6}$ & 72 \\
\bottomrule
 \end{tabular}
\label{tab:rayleigh}
\end{table}

In all the simulations presented in this study, the Ekman number is kept fixed
to a moderate value of $\text{E}=10^{-4}$, which
allows to study a large number of cases. The Prandtl number
is set to 1 and the magnetic Prandtl number to 2. Following \cite{Jones09}
and our previous hydrodynamical models \citep{Gastine12}, we adopt a polytropic
index $m=2$ for the reference state. We consider two different
aspect ratios: thick shells with $\eta=0.2$ and thin shells with $\eta=0.6$.
Dynamo models in thin shells rely on the same setup as our previous
hydrodynamical study \citep{Gastine12}.
Gravity is adapted to better reflect the different shells, i.e. we use linear
gravity in the thick shells ($g_1=1,\,g_2=0$) and $g\propto 1/r^2$ in the thin shells
($g_1=0,\,g_2=1$). Both cases also differ in the flow boundary conditions:
thin shell cases assume stress-free conditions at both boundaries, thick shell
cases assume a rigid inner boundary.

Being mostly interested in the effects of the density
stratification, we have performed numerical simulations with different density
contrasts spanning the range from $N_\rho=0$ (i.e. Boussinesq) to $N_\rho=3$
(i.e. $\rho_\text{bot}/\rho_\text{top}\simeq 20$). For each density
stratification, we vary the Rayleigh number from simulations close to onset of
dynamo action to 10-50 times the critical Rayleigh number $\text{Ra}_c$ for the
onset of convection. $\text{Ra}_c$ varies with $N_\rho$ and we list respective
values for the cases explored here in Tab.~\ref{tab:rayleigh}. These have
been determined with the anelastic linear
stability analysis code developed by \cite{Jones09a}.

As will be shown below, we find several cases of bistability where
a strong dipole-dominated and a weak multipolar solution coexist at identical
parameters. The starting solution for our time integration
then determines which of the two solutions the simulation will assume.
Most of the numerical simulations therefore have been initiated with both a
strong dipolar magnetic field (with $\Lambda \sim 1$) and a weak
multipolar magnetic field (model names
end with a  'd' or a 'm' accordingly in Tab.~\ref{tab:results}). Altogether,
more than 100 simulations have been computed, each running several magnetic
diffusion time to ensure that a stable magnetic configuration has been reached.
Table~\ref{tab:results} lists the input parameters of all cases
along with key solution properties.

\section{Results}
\label{sec:results}

\subsection{Dynamo regimes}

\begin{figure}
 \centering
 \includegraphics[width=9cm]{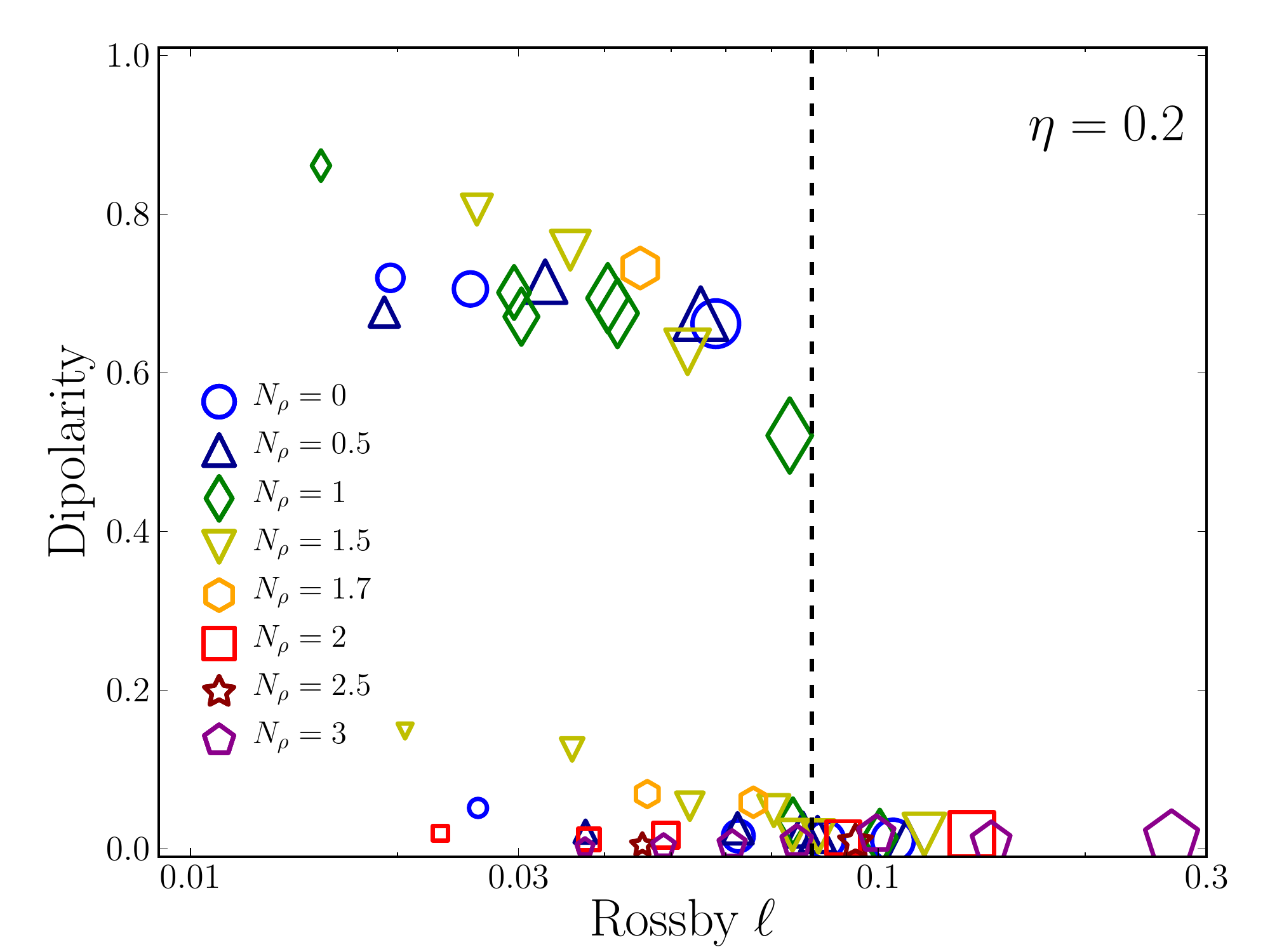}
 \includegraphics[width=9cm]{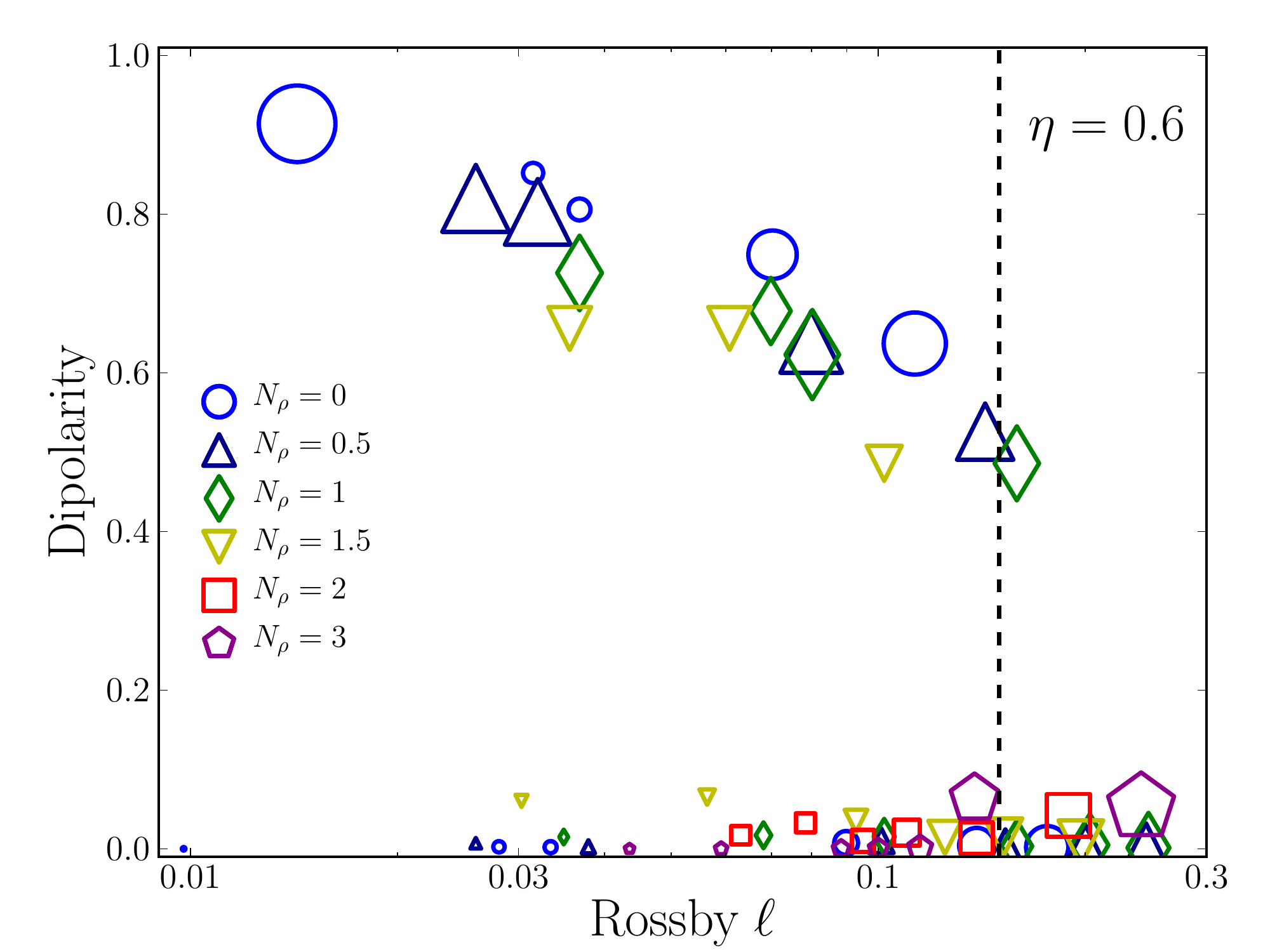}
 \caption{Relative dipole strength plotted against the local Rossby number
$\text{Ro}_\ell$ for two different aspect ratios:
$\eta=0.2$ (upper panel) and $\eta=0.6$ (lower panel). Each type of
symbol is associated to a given density stratification (from Boussinesq, i.e.
$N_\rho=0$ to stratified simulations corresponding to $N_\rho=3$). The size of
the symbol has been chosen according to the value of the modified Elsasser
number (Eq.~\ref{eq:elsloc}). Vertical lines are tentative transitions between
dipolar and multipolar dynamos.}
 \label{fig:dipol}
\end{figure}

Figure~\ref{fig:dipol} shows how the dipolarity $f_\text{dip}$  depends
on the local Rossby number in the thick shell (top) and the thin shell
(bottom) simulations. In both geometries we find two distinct branches:
The upper branch corresponds to the dipolar regime  at $f_\text{dip} > 0.5$
with strong magnetic fields at modified Elsasser number between
 $\Lambda_\ell= 0.1$ and $\Lambda_\ell =1$. This indicates a strong Lorentz force so that
these dynamos likely operate in the so-called magnetostrophic regime
where Coriolis forces, pressure gradients, buoyancy and Lorentz forces
contribute to the first order Navier-Stokes equation.
In contrast, the models belonging to the lower branch have a multipolar field
geometry at $f_\text{dip} < 0.2$ and much weaker magnetic
field with $\Lambda_\ell \lesssim 5\times 10^{-2}$. Since the Lorentz force
may thus not enter the first order Navier-Stokes equation
these cases are likely geostrophic.

\begin{figure}
 \centering
 \includegraphics[width=9cm]{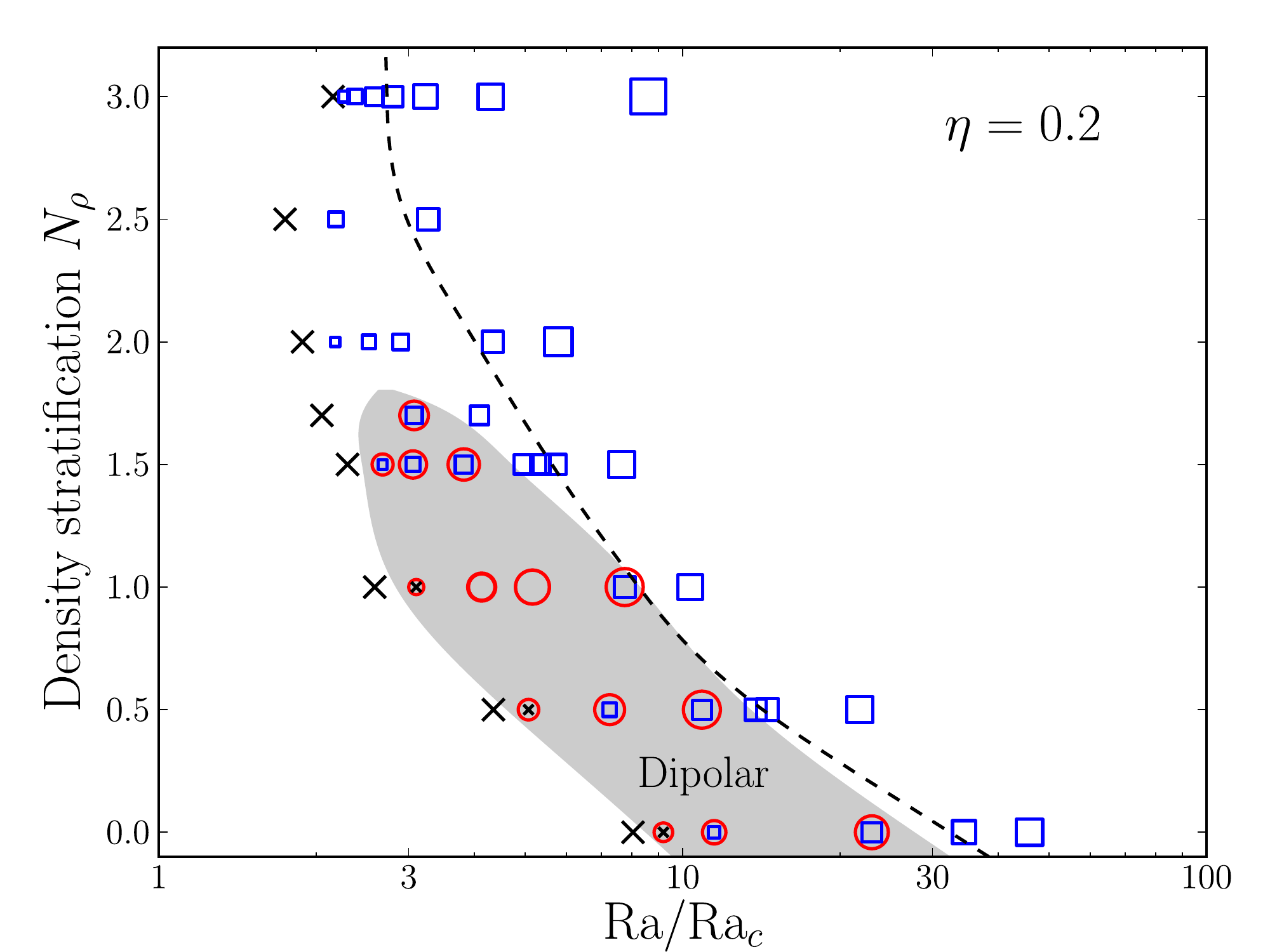}
 \includegraphics[width=9cm]{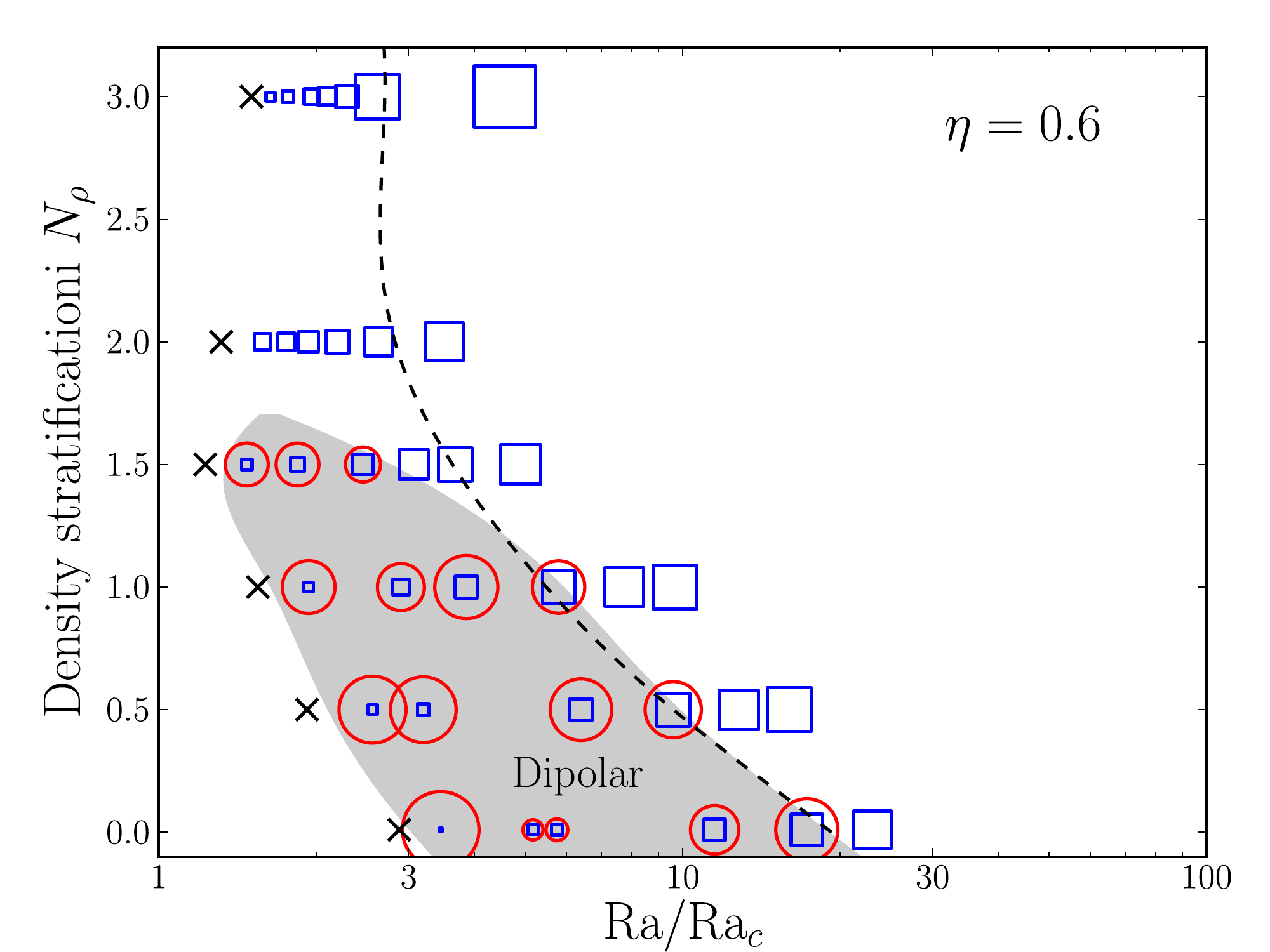}
 \caption{Regime diagram for dynamo simulations as a function of
supercriticality and density stratification for two different aspect ratios:
$\eta=0.2$ (upper panel) and $\eta=0.6$ (lower panel). Red circles correspond to
dipolar dynamos, blue squares to multipolar dynamos and black crosses are
decaying dynamos. The size of the symbols depend on the value of the modified
Elsasser number (Eq.~\ref{eq:elsloc}). The grey-shaded area highlights the
dipolar region, while the dashed lines correspond to the critical
$\text{Ro}_{\ell c}$ values that mark the limit between dipolar and multipolar
dynamos (see the vertical lines in Fig.~\ref{fig:dipol}).}
 \label{fig:crit}
\end{figure}

For both aspect ratios, the dipolar branch is bounded by a critical local
Rossby number $\text{Ro}_{\ell c}$ beyond which all cases are multipolar.
In the thin shell with $\eta=0.6$, the value $\text{Ro}_{\ell c}\simeq 0.15$
is slightly larger than the value of $\text{Ro}_{\ell c}\simeq 0.12$ predicted
by \cite{Christensen06} in their geodynamo models with $\eta=0.35$. In the 
thick shell with $\eta=0.2$,
it is smaller at $\text{Ro}_{\ell c}\simeq 0.08$. The critical
local Rossby number thus seems to increase with the inner core size.
A similar dependence on the aspect ratio has been already found by 
\cite{Aubert09}.
In contrast to the previous Boussinesq studies that used rigid boundary
conditions \citep[e.g.][]{Christensen06,Aubert09}, the multipolar
branch also extends here below the critical $\text{Ro}_{\ell c}$ where
the dipolar and the multipolar branch now coexist. This multipolar dynamo
branch for $\text{Ro}_{\ell} < 0.1$ has also been found in the Boussinesq
models of \cite{Schrinner12} with stress-free boundary conditions.
We further discuss the bistability phenomenon in section~\ref{sec:bi}.

Figure \ref{fig:crit} illustrates how the field morphology depends on
the stratification and on the supercriticality $\text{Ra}/\text{Ra}_c$.
The value $N_\rho\simeq 1.8$ marks an important regime boundary:
below this value most cases are bistable (nested circles and squares),
while above this value only multipolar cases remain (squares only).
The reason for this boundary will be considered in section~\ref{sec:strati}.
The dashed black lines in Fig.~\ref{fig:crit} mark the $\text{Ro}_{\ell c}$
boundary which moves
to lower supercriticalities for increasing density stratifications.
Larger stratifications promote larger flow amplitudes as well
as smaller lengthscales since convection is progressively concentrated in
a narrowing region close to the outer boundary \citep[see
Table~\ref{tab:rayleigh}, Fig.~\ref{fig:vortz} and][]{Jones09,Gastine12}. Both
effects lead to larger $\text{Ro}_\ell$ for a given supercriticality.

For $N_\rho< 1.8$ the boundary for onset of dynamo action
also seems to move to smaller supercriticalities when $N_\rho$ increases.
The reason is a mixture of a decreasing critical magnetic Reynolds number
and an increasing flow amplitude for a given supercriticality.
Beyond $N_\rho=1.8$, the critical magnetic Reynolds number once
more increases and the onset of dynamo action moves to larger
supercriticalities. The onset of dynamo action and the $\text{Ro}_{\ell c}$
boundary are less affected by the stratification for the thinner shell.
One reason may be that the flow lengthscale is already quite small
for thinner shells even in the Boussinesq case \citep[see
Tab.~\ref{tab:rayleigh} and][]{Shamali04}. The differences in the gravity
profile and the flow boundary condition may also play a role here.

The dipolar branch reaches down to lower local Rossby numbers than the
multipolar branch (see Fig.~\ref{fig:dipol}). This is where we find
examples of subcritical dynamo action (models 1d, 6d and 12d in
Tab.~\ref{tab:results}), where the dynamo is only
successful when started with a sizable magnetic field strength \citep{Morin09}.
Simulations started with a weak field do not have the option to
run to the multipolar branch and the field simply decays away.

\subsection{Bistability}
\label{sec:bi}

\begin{figure}
 \centering
 \includegraphics[width=9cm]{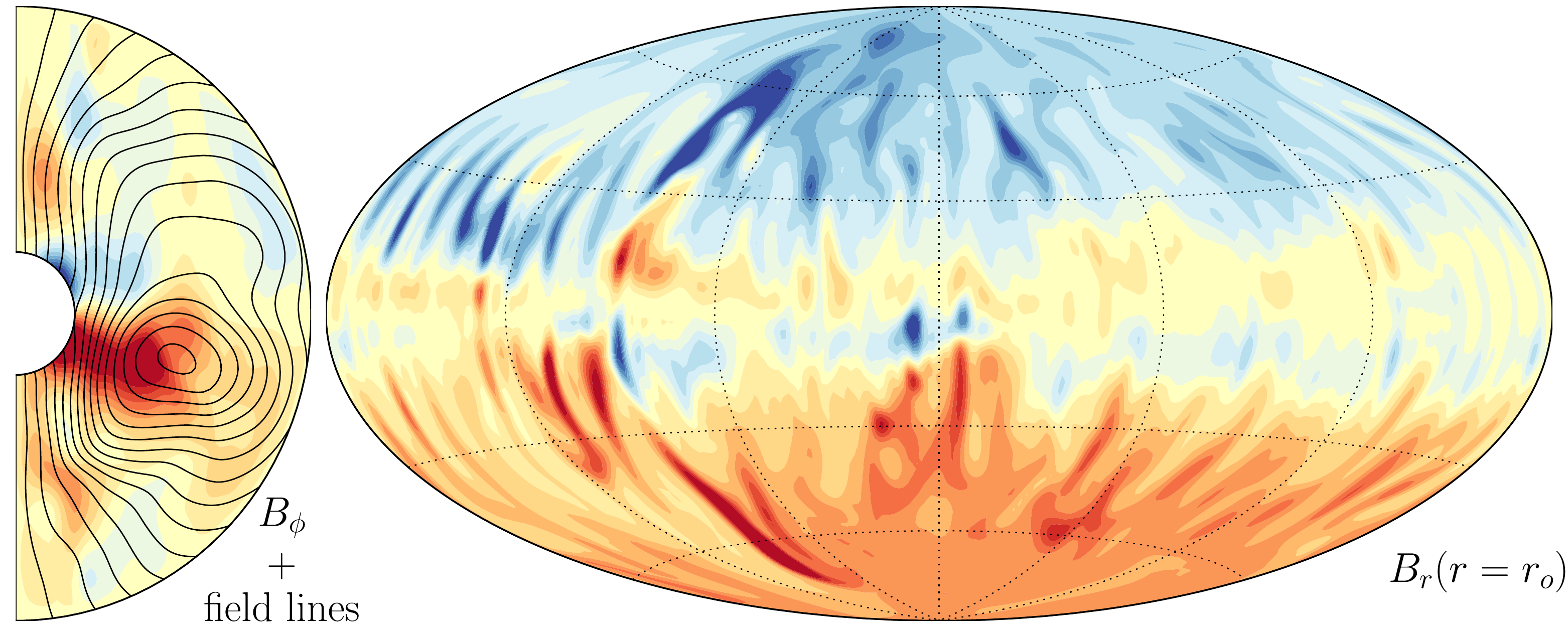}
 \includegraphics[width=9cm]{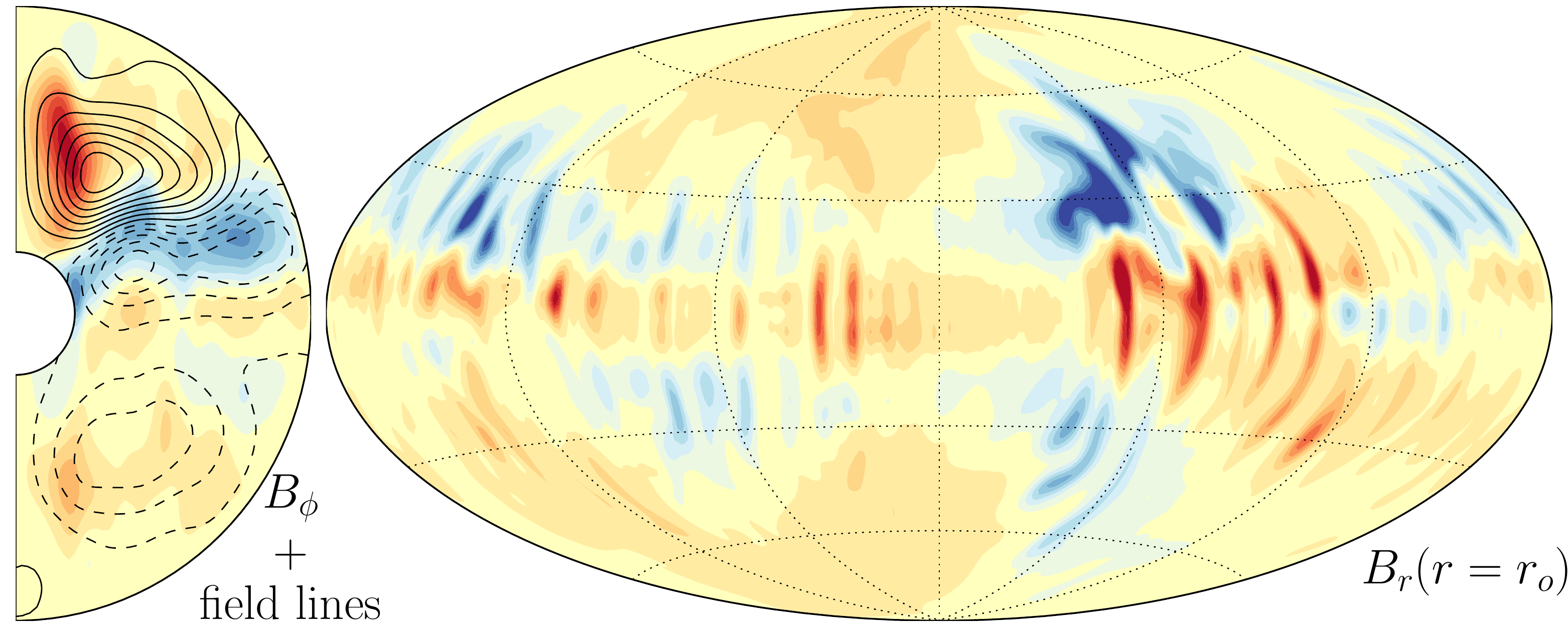}
\caption{Left panels: iso-contours of axisymmetric azimuthal field and poloidal
field lines. Right panels: surface radial field $B_r(r=r_o)$. From a model with
$\eta=0.2$, $N_\rho=1.7$, $\text{Ra}=9\times 10^{6}$ initiated with a dipolar
magnetic field (upper panels) and a multipolar one (lower panels). Red (blue)
correspond to positive (negative) values.}
\label{fig:bistab}
\end{figure}

\begin{figure}
 \centering
 \includegraphics[width=9cm]{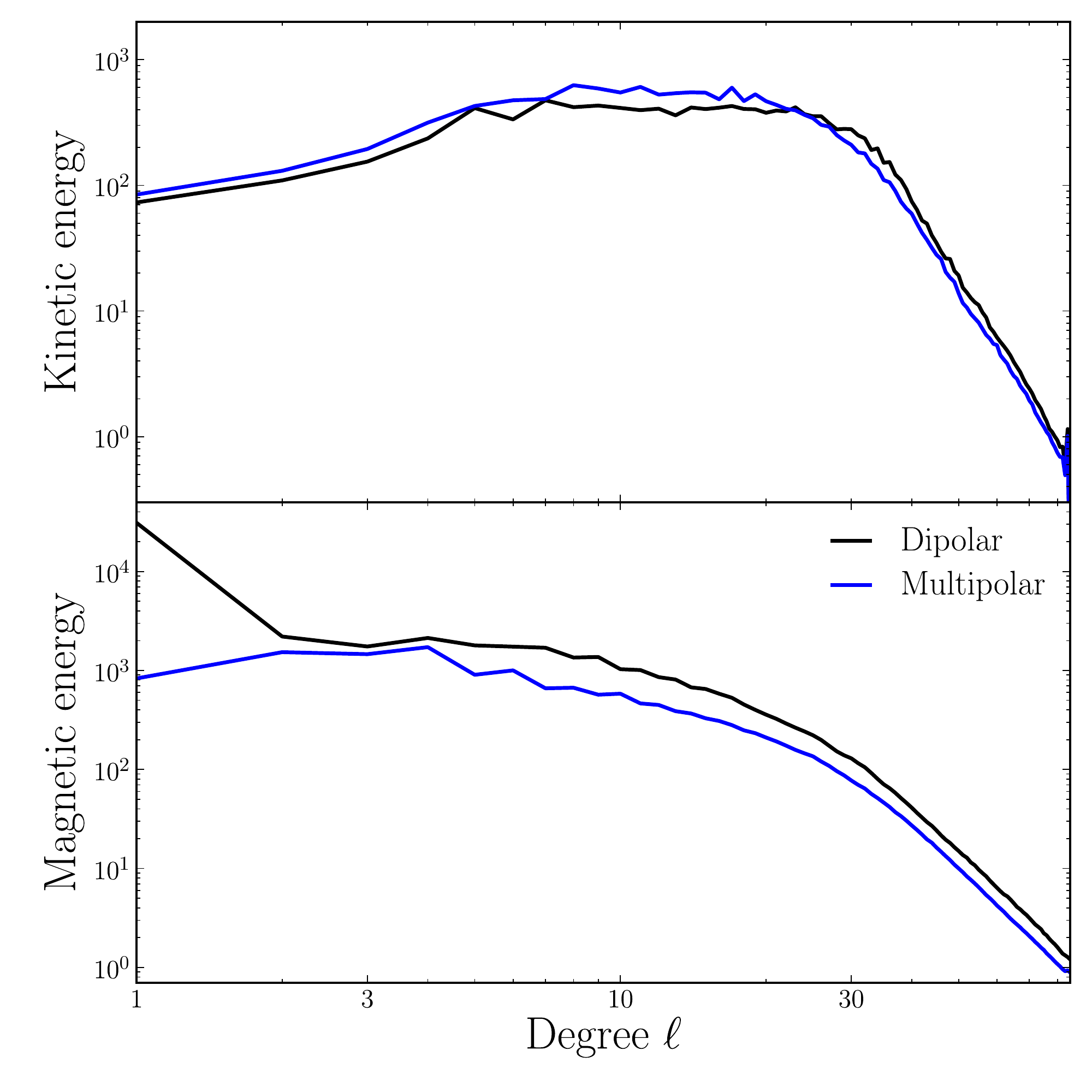}
\caption{Time-averaged kinetic and magnetic energy spectra plotted against the
spherical harmonic degree $\ell$. From a model with
$\eta=0.2$, $N_\rho=1.7$, $\text{Ra}=9\times 10^{6}$ initiated with a dipolar
magnetic field (black lines) and a multipolar one (blue lines), corresponding
to Fig.~\ref{fig:bistab}.}
\label{fig:spec}
\end{figure}

As illustrated on Fig.~\ref{fig:crit}, most of our models
at $N_\rho<1.8$ show bistability. Exceptions are the subcritical thick shell cases
and models 13d/m and 14d/m (see Tab.~\ref{tab:results}), where even small
initial magnetic fields developed to strong dipoles.
The top panel of Fig.~\ref{fig:bistab} shows the dipole-dominated solution with,
however, stronger dynamo action in the southern hemisphere.
The lower panel of Fig.~\ref{fig:bistab} illustrates the weak multipolar
solution, which has pronounced equatorially symmetric and non-axisymmetric
components (this is a feature also visible in the more stratified cases
shown in the following Figs.~\ref{fig:brsurfEta02}-\ref{fig:brsurfEta06}).
The dipolar solution has a three times larger modified Elsasser number.  
Generally, $\Lambda_\ell$ is twice to ten times larger
in the dipole-dominated than in the multipolar counterpart of bistable cases
(see Fig.~\ref{fig:dipol} and Tab.~\ref{tab:results} for further details). 
This suggests that the two dynamo branches are also characterised by different
force balances, at least in the bistability region (i.e.
$\text{Ro}_\ell < 0.1$): the dipolar branch is magnetostrophic, while the
multipolar one is more likely to be geostrophic. Beyond $\text{Ro}_\ell > 0.1$,
however, the multipolar dynamos can be strong enough to yield a magnetostrophic
force balance due to the larger Rayleigh numbers \citep{Brun05}.

Figure~\ref{fig:spec} shows the corresponding time-averaged kinetic and
magnetic spectra for the two bistable solutions displayed in
Fig.~\ref{fig:bistab}. The two kinetic energy spectra nearly coincide,
emphasising the similarities of the flow in the two solutions. We observe a
broad plateau for degrees $4 \leq \ell \leq 30$ that
reflects the convective driving (for this model, the critical
azimuthal wavenumber at onset of convection is $m=6$, see
Tab.~\ref{tab:rayleigh}). The magnetic spectra show the differences already
visible in the surface magnetic fields displayed in Fig.~\ref{fig:bistab}. The
amplitude of the dipolar component is more than one order of magnitude lower,
higher-order components drop by roughly $50\%$. A mild maximum for  $\ell=4$
corresponds to the large-scale contribution visible in Fig.~\ref{fig:bistab}.
For spherical harmonic degrees $\ell > 30$, the spectra follow some
clear power-law behaviour. Both magnetic and kinetic
energy spectra have a steep decrease (between $\ell^{-4}$ and $\ell^{-5}$), very
similar to previous stellar dynamo models \citep[e.g.][]{Brun05,Browning08} or
quasi-geostrophic kinematic dynamos \citep[e.g.][]{Schaeffer06}.

The coexistence of a dipolar and a multipolar branch
has already been reported for Boussinesq models with
stress-free \citep{Busse06,Schrinner12} or mixed mechanical boundary
conditions \citep{Sasaki11}. For rigid boundary conditions only one
such case has been found by \cite{Christensen06}.
Stress-free boundaries allow stronger zonal winds to develop
which seem to play a key role here.
These geostrophic flows (constant on coaxial cylinders) are maintained by
Reynolds stresses, i.e. by the correlation between the zonal $u_\phi$ and
the cylindrically radial $u_s$ velocity components \citep[e.g.][]{Gastine12}.
For stress-free boundary conditions their amplitude is limited by the
weak bulk viscosity. When rigid boundary conditions are employed, the
much stronger boundary friction severely brakes these zonal
winds.

\begin{figure}
 \centering
 \includegraphics[width=9cm]{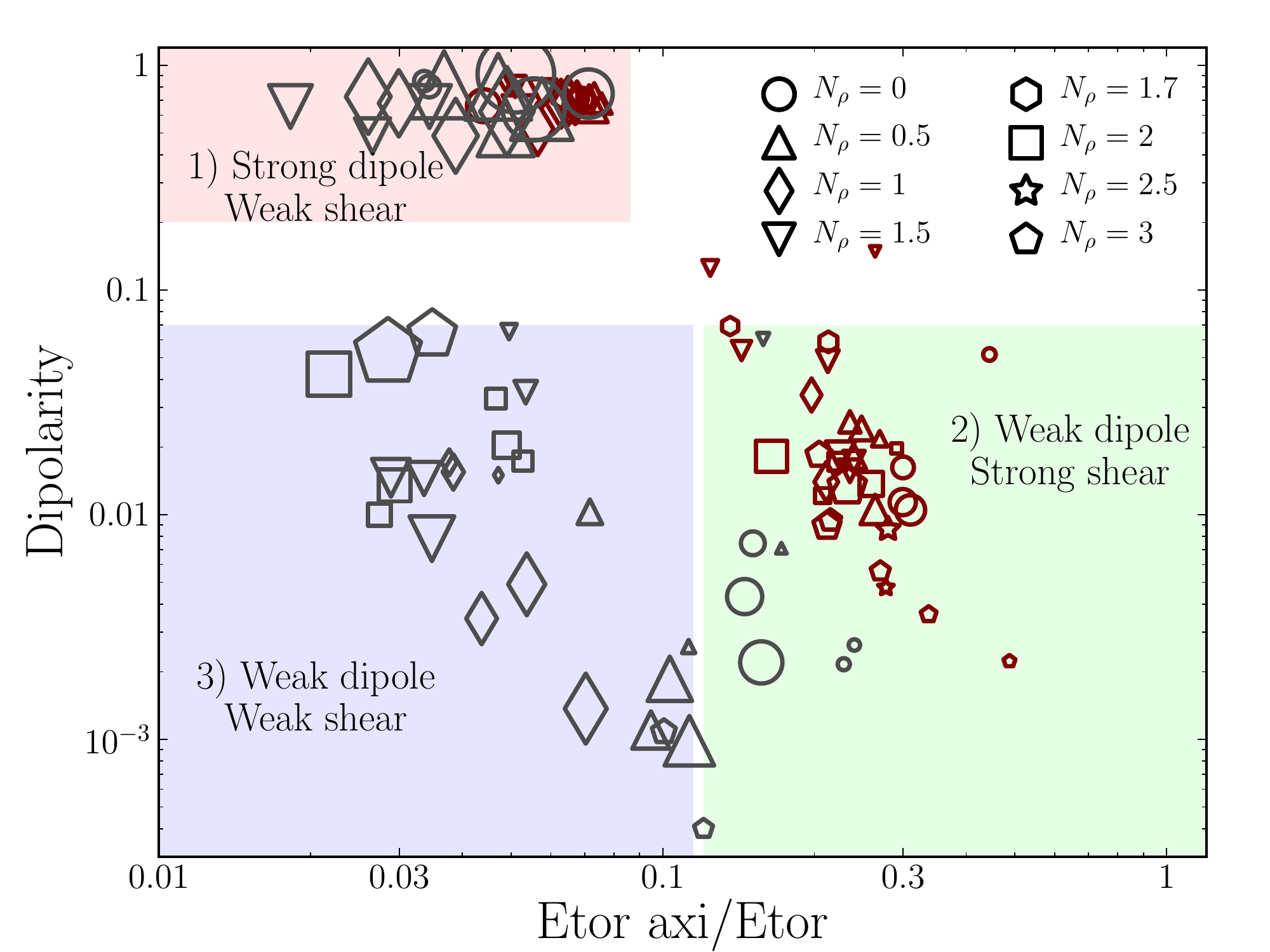}
 \caption{Relative dipole strength plotted against the ratio of axisymmetric
toroidal kinetic energy divided by toroidal energy.
Red (grey) symbols correspond to simulations in
thick (thin) shells ($\eta=0.2$ and $\eta=0.6$, respectively). Each type of
symbol is associated with a given density stratification (from Boussinesq, i.e.
$N_\rho=0$ to $N_\rho=3$). The sizes of the symbols have been chosen according
to the values of the modified Elsasser number (Eq.~\ref{eq:elsloc}).}
 \label{fig:shear}
\end{figure}

Figure~\ref{fig:shear} displays the
dipolarity of the surface field against the ratio of axisymmetric
toroidal to total toroidal kinetic energy, which
is a good proxy of the relative amplitude of zonal winds.
Strongly dipolar solutions cluster in the upper left corner where zonal
winds are weak. For strong zonal winds only weak multipolar solutions can be found.
For bistable cases, one solution belongs to the first type while the
other one belongs to the second category.
There clearly is a competition between strong zonal winds and
strong dipolar magnetic fields.
The third type of multipolar and sometimes strong solutions with mostly
weak zonal winds are mainly strongly stratified
cases that we will discuss in section \ref{sec:strati}. The strong multipolar 
fields can have a strong impact on the zonal flows,
in a similar way to the stellar models of \cite{Brun05}.

\begin{figure}
 \centering
 \includegraphics[width=4.4cm]{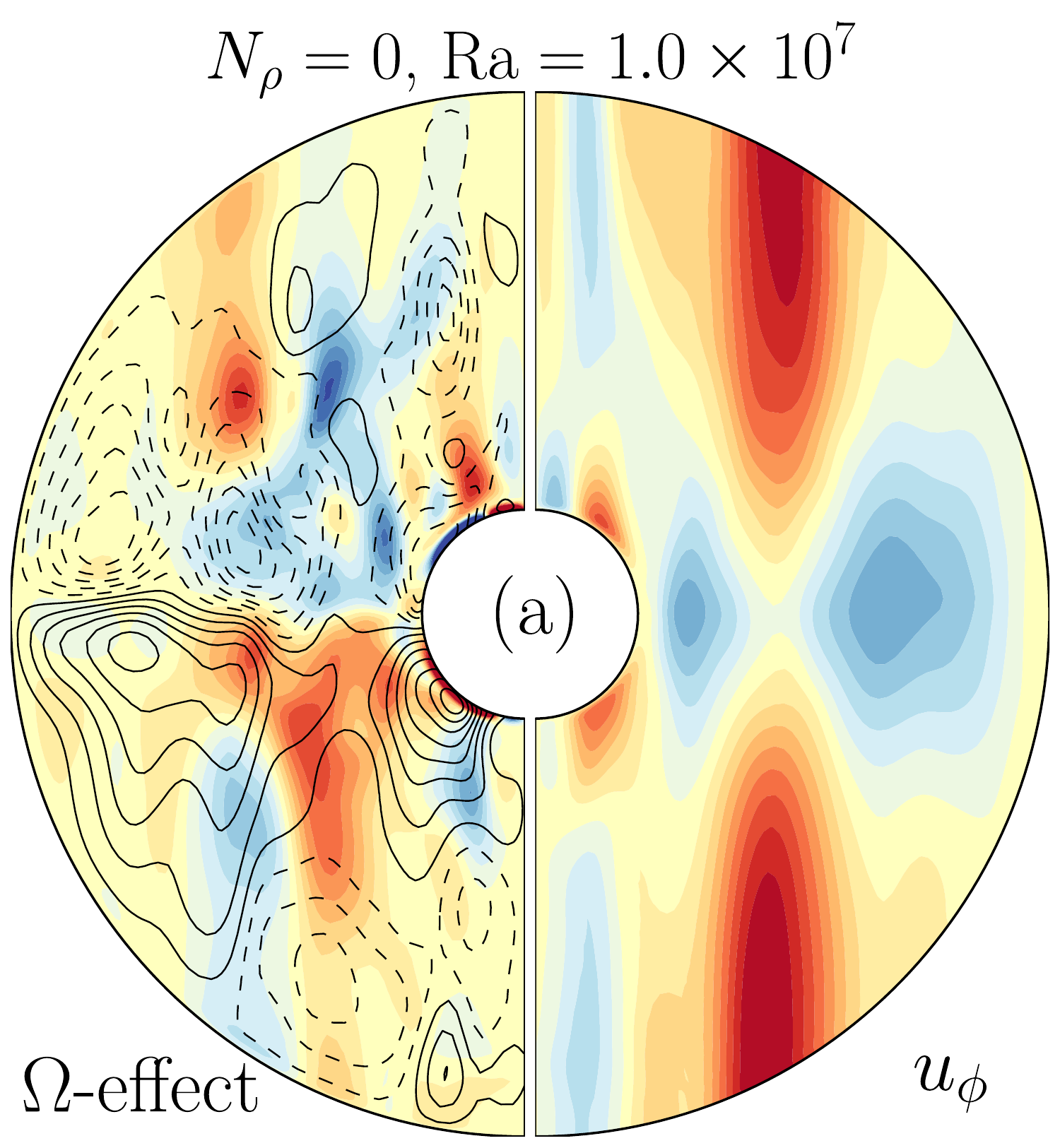}
 \includegraphics[width=4.4cm]{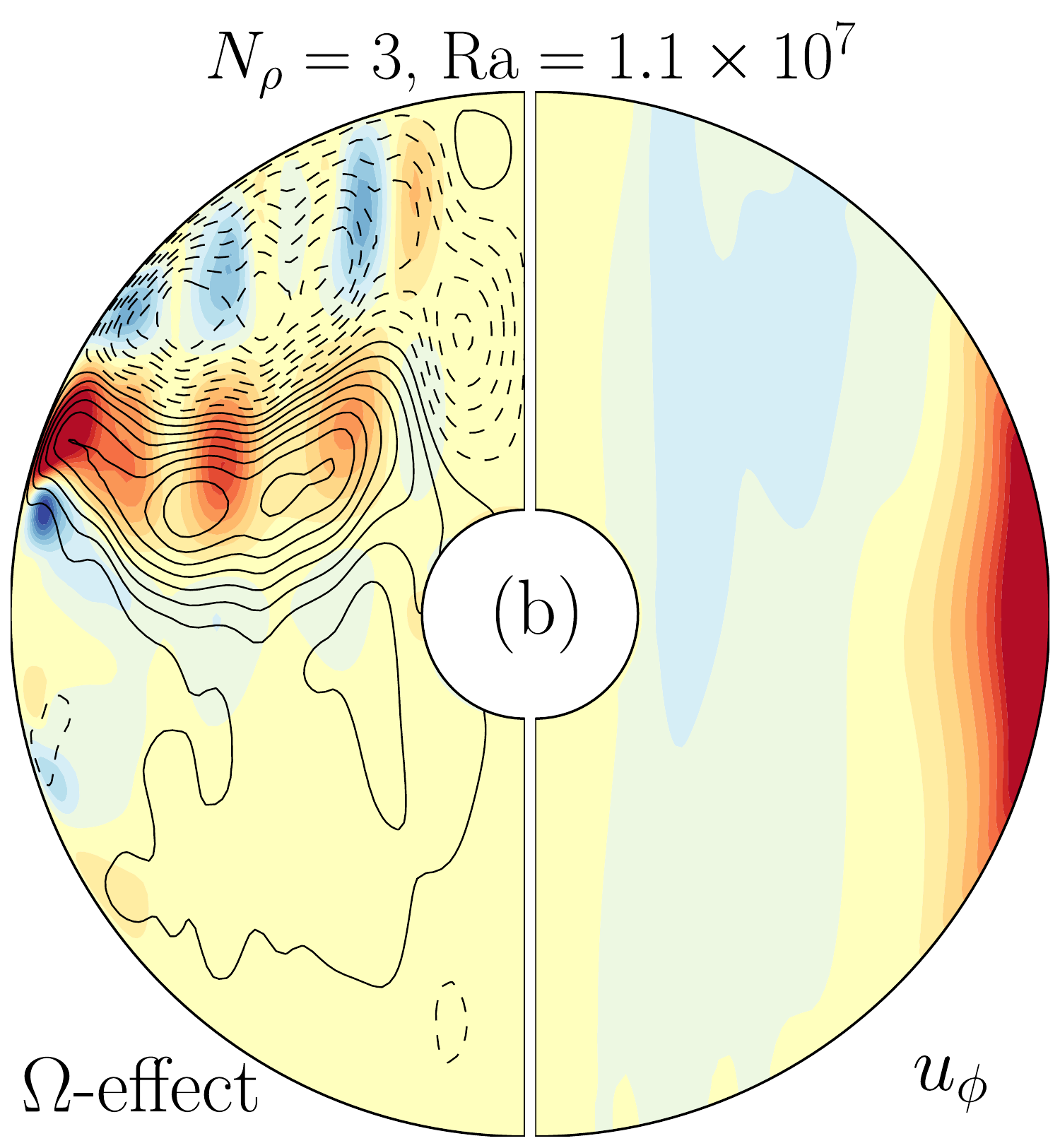}
 \includegraphics[width=4.4cm]{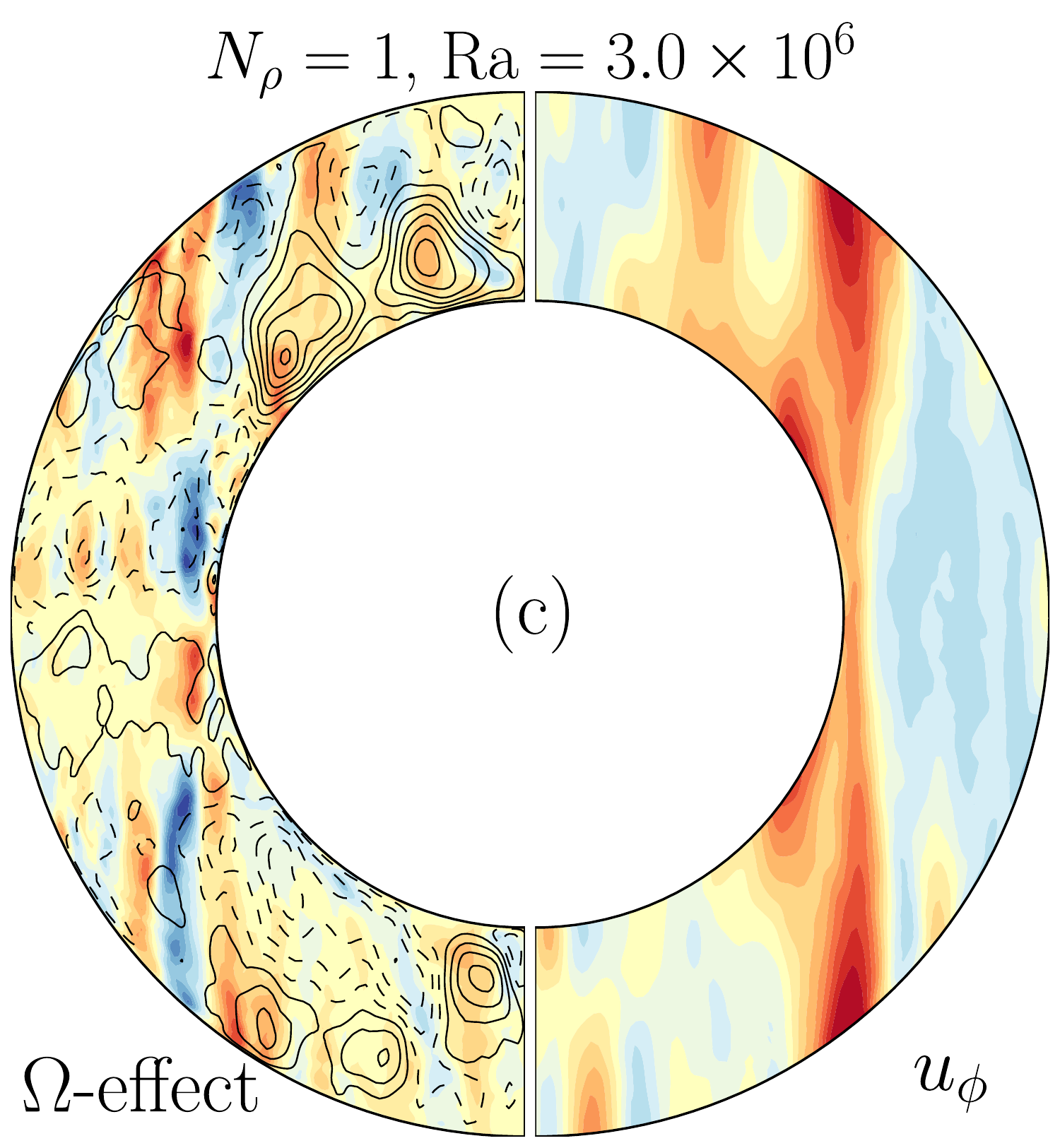}
 \includegraphics[width=4.4cm]{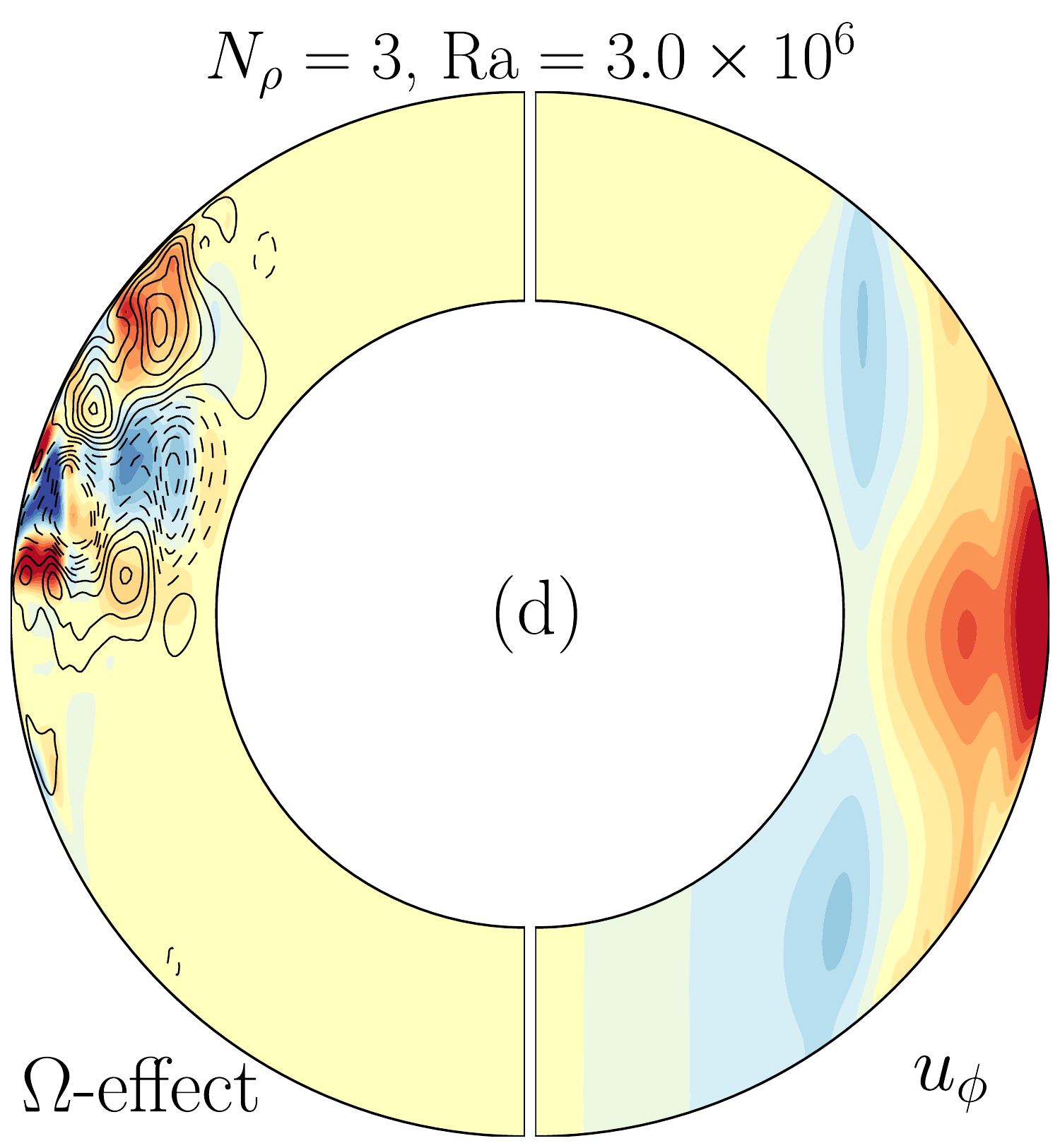}
\caption{Role of the $\Omega$-effect in the production of the toroidal field
for two selected simulations in thick shells (upper panels) and in thin shells
(lower panel). For each simulation, the $\Omega$-effect (color levels) and the
axisymmetric azimuthal field (solid and dashed lines) are
displayed on the left part, while the axisymmetric zonal flow is given on the
right part. Red (blue) correspond to positive (negative) values.}
 \label{fig:omeffect}
\end{figure}

The stronger zonal winds in the multipolar cases go along with
a change in dynamo mechanism which is illustrated in Fig.~\ref{fig:omeffect}.
In the dipolar cases (left panels), zonal flows are weak and
mainly driven by thermal wind effects.
The stronger Lorentz force can balance the Coriolis force and
allows for these strongly non-geostrophic motions \citep{Aubert05}.
The $\Omega$-effect, i.e. the production of axisymmetric toroidal magnetic
field by zonal wind shear, plays only a secondary role
(left half of left panels in Fig.~\ref{fig:omeffect}) so that the
dynamo is of the $\alpha^2$ type in the mean field nomenclature
\citep{Olson99}, similar to the planetary dynamo models
by \cite{Christensen06} or mean-field models of fully
convective stars \citep{Chabrier06}.
In the multipolar cases (right panels in Fig.~\ref{fig:omeffect}),
significant zonal winds driven by Reynolds stresses develop close to the outer
boundary and the associated $\Omega$-effect plays an important role for
toroidal magnetic field production. Dynamos on the multipolar branch
(for $N_\rho<2$) are thus of the $\alpha\Omega$ or $\alpha^2\Omega$
type. Many of these solutions show distinct oscillations connected
to Parker dynamo waves that we discuss further in section \ref{sec:parker}.

\subsection{Influence of stratification}
\label{sec:strati}

\begin{figure*}
 \centering
 \includegraphics[width=\textwidth]{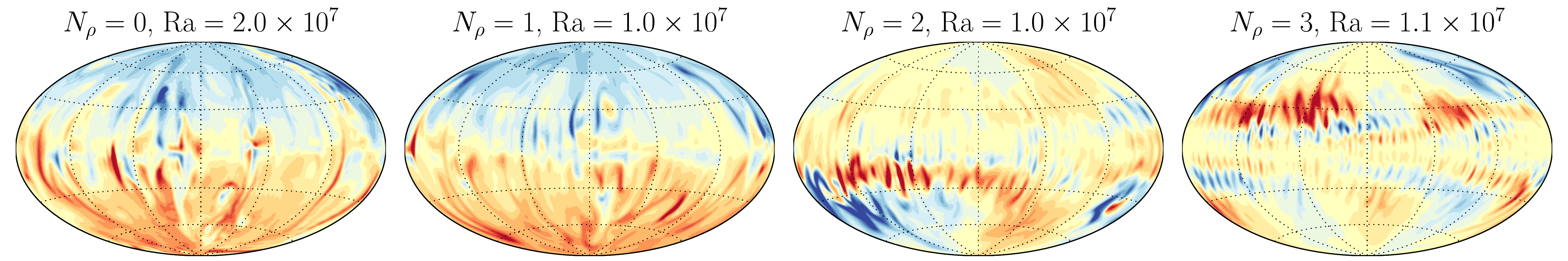}
 \includegraphics[width=\textwidth]{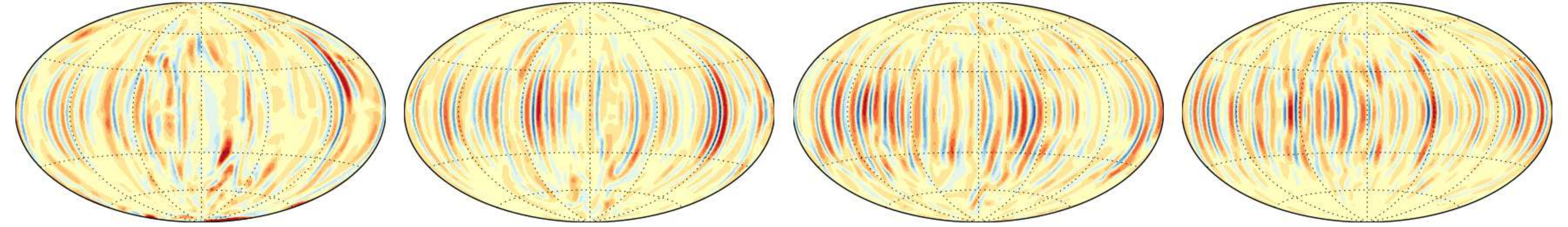}
\caption{Surface radial field $B_r(r=r_o)$ (upper panels) and radial
velocity close to the outer boundary $u_r(r=0.9\,r_o)$ (lower panels) for four
numerical simulations in
thick shells (i.e. $\eta=0.2$) with similar local Rossby number
$\text{Ro}_\ell \simeq 0.04-0.05$ (see Tab.~\ref{tab:results}). Red (blue)
correspond to positive (negative) values.}
\label{fig:brsurfEta02}
\end{figure*}

\begin{figure*}
 \centering
 \includegraphics[width=\textwidth]{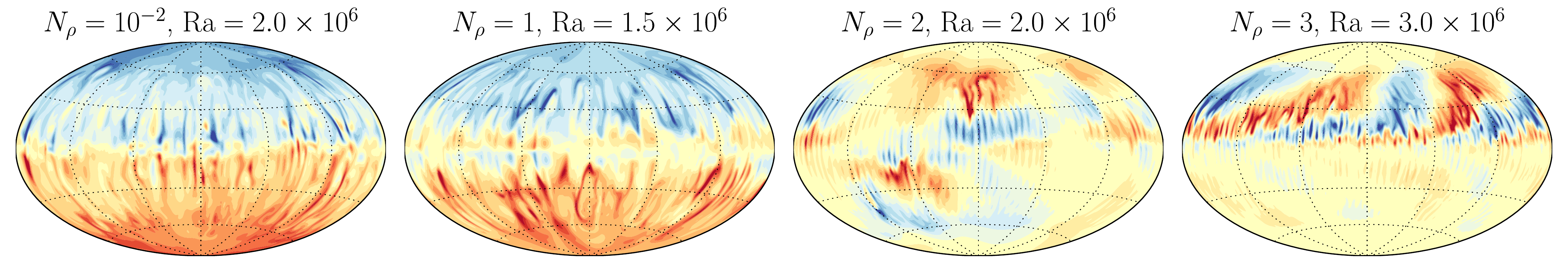}
 \includegraphics[width=\textwidth]{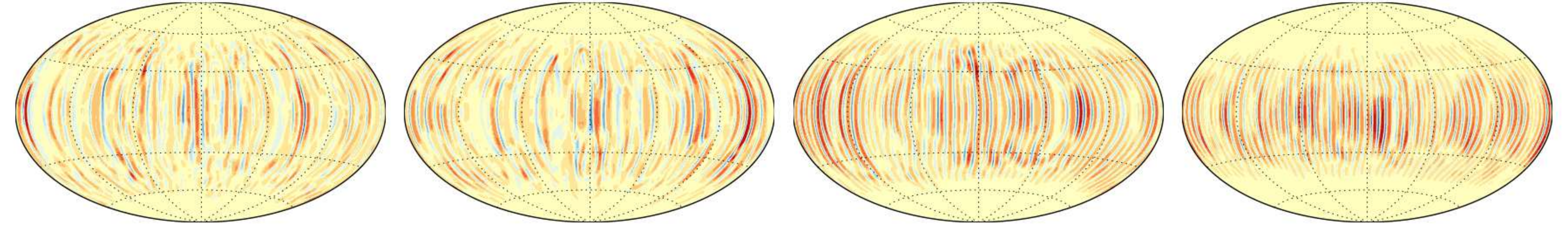}
\caption{Surface radial field $B_r(r=r_o)$ (upper panels)  and radial
velocity  close to the outer boundary $u_r(r=0.9\,r_o)$ (lower panels) for
four numerical simulations in thin shells (i.e. $\eta=0.6$) with similar local
Rossby number
$\text{Ro}_\ell \simeq 0.07-0.08$ (see Tab.~\ref{tab:results}). Red (blue)
correspond to positive (negative) values.}
\label{fig:brsurfEta06}
\end{figure*}

\begin{figure*}
 \centering
 \includegraphics[width=\textwidth]{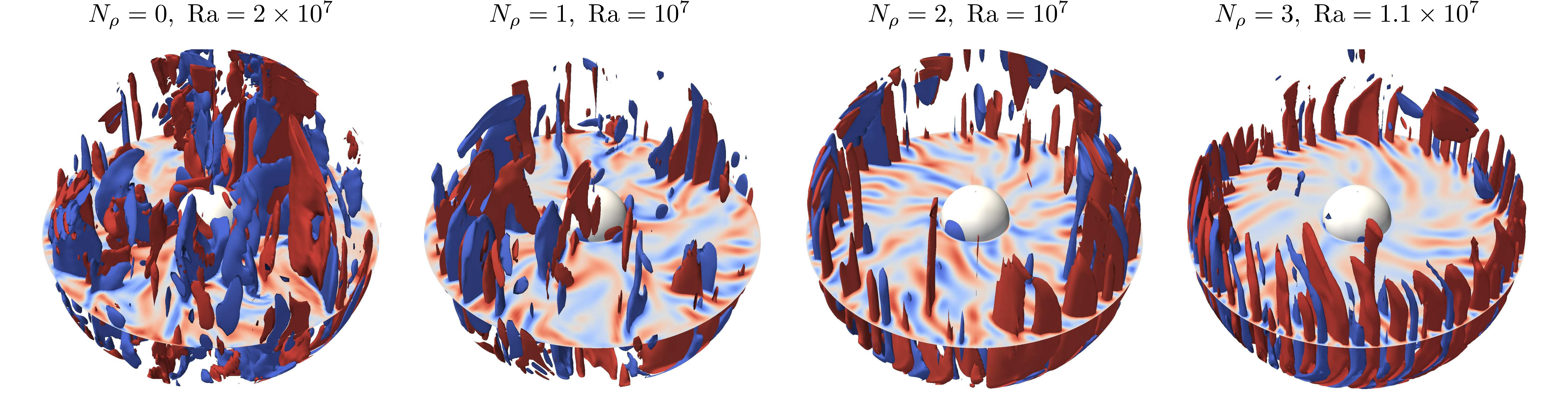}
 \includegraphics[width=\textwidth]{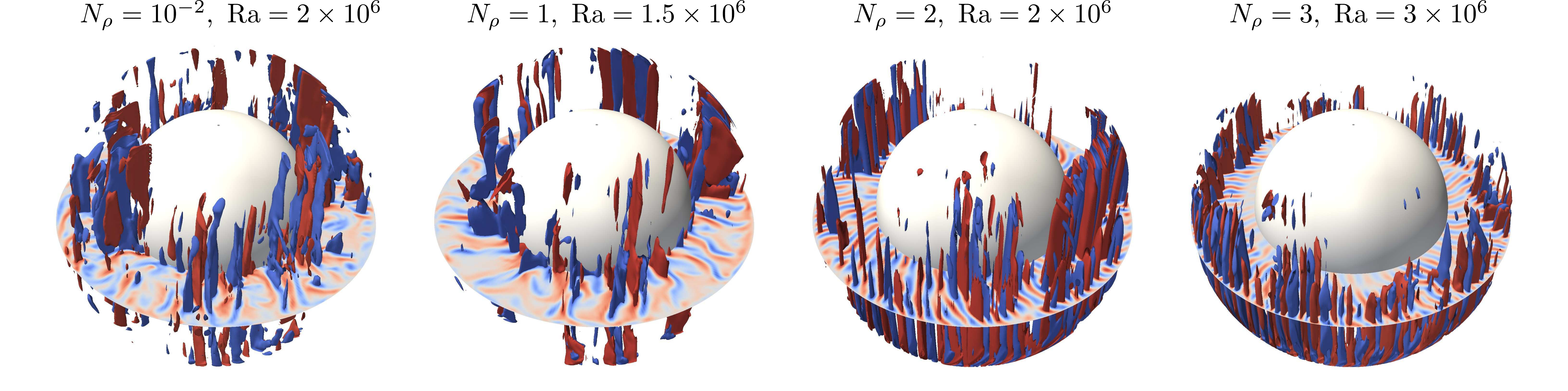}
\caption{Isocontours and equatorial cut of vorticity along the axis of rotation
$\omega_z=(\vec{\nabla}\times\vec{u})_z$ for different density stratification in
thick shells ($\eta=0.2$, upper panels) and in thin shells ($\eta=0.6$, lower
panels). These models correspond to the cases displayed in
Figs.~\ref{fig:brsurfEta02}-\ref{fig:brsurfEta06}. Positive (negative) values
are rendered in red (blue). For each model, the white sphere corresponds to the
inner radius.}
\label{fig:vortz}
\end{figure*}

For $N_\rho>1.8$ only the multipolar branch remains.
Figures~\ref{fig:brsurfEta02} and \ref{fig:brsurfEta06} illustrate how
the convective flow and the surface magnetic field evolve on increasing the
density stratification for simulations with very similar local Rossby numbers.
All these models have been initiated with a strong dipolar field. For both
aspect ratios, the weakly stratified cases (i.e. Boussinesq and $N_\rho=1$) have
a dipole-dominated magnetic field, while the stronger stratified cases (i.e. 
$N_\rho=2$ and $N_\rho=3$) have multipolar fields with significant
non-axisymmetric contributions. As a consequence of the background
density contrast, the typical lengthscale of convection gradually decreases
with $N_\rho$ \citep[see also Tab.~\ref{tab:rayleigh} and][]{Gastine12}.
While the imprints of these smaller convective scale are visible in the surface
magnetic fields, the dynamos are always dominated by larger scale features even
in the strongly stratified cases. For example, for the two models at $N_\rho=3$,
a clear wave number $m=2$ signature emerges similar to the patterns already
observed in the Boussinesq models of \cite{Goudard08}.
This magnetic mode can only be clearly identified for moderate magnetic
Reynolds numbers $\text{Rm} < 200$ close to the onset of dynamo action.
For larger Rm, smaller scale more complex features once more take over.

Some models also show a stronger concentration of the magnetic field in one
hemisphere. Similar effects can be found for some multipolar simulations
at low stratifications (see section \ref{sec:parker}).
\cite{Grote00} and \cite{Busse06} report that
hemispherical dynamo action is typical for simulations with stress-free boundaries
and Prandtl numbers around unity.

The collapse of the dipolar branch for $N_\rho\gtrsim 2$ is caused by the
strong concentration of convective features close to the outer boundary where
density decreases most drastically.
Figure~\ref{fig:vortz} illustrates the evolution of the convective
columns when $N_\rho$ increases while the local Rossby
numbers remains similar
(see Figs.~\ref{fig:brsurfEta02}-\ref{fig:brsurfEta06} for the
corresponding magnetic fields and radial flow structures).
As it has already been demonstrated by non-magnetic simulations
\citep{Jones09a,Gastine12}, convection first develops close to the inner
boundary for Boussinesq and weakly stratified models ($N_\rho \leq 1$).
When increasing $N_\rho$, however, the convective columns gradually move
outward and become confined to a thin region close to the equator
($N_\rho \geq 2$). This goes along with a decrease of
the typical lengthscale (see also
Tab.~\ref{tab:rayleigh} for the critical azimuthal wave numbers at onset).
According to \cite{Gastine12}, this is a consequence of the fact that
buoyancy and thus the effective local Rayleigh number becomes much larger
at the outer boundary than in the interior for strongly stratified models.

\begin{figure}
 \centering
 \includegraphics[width=9cm]{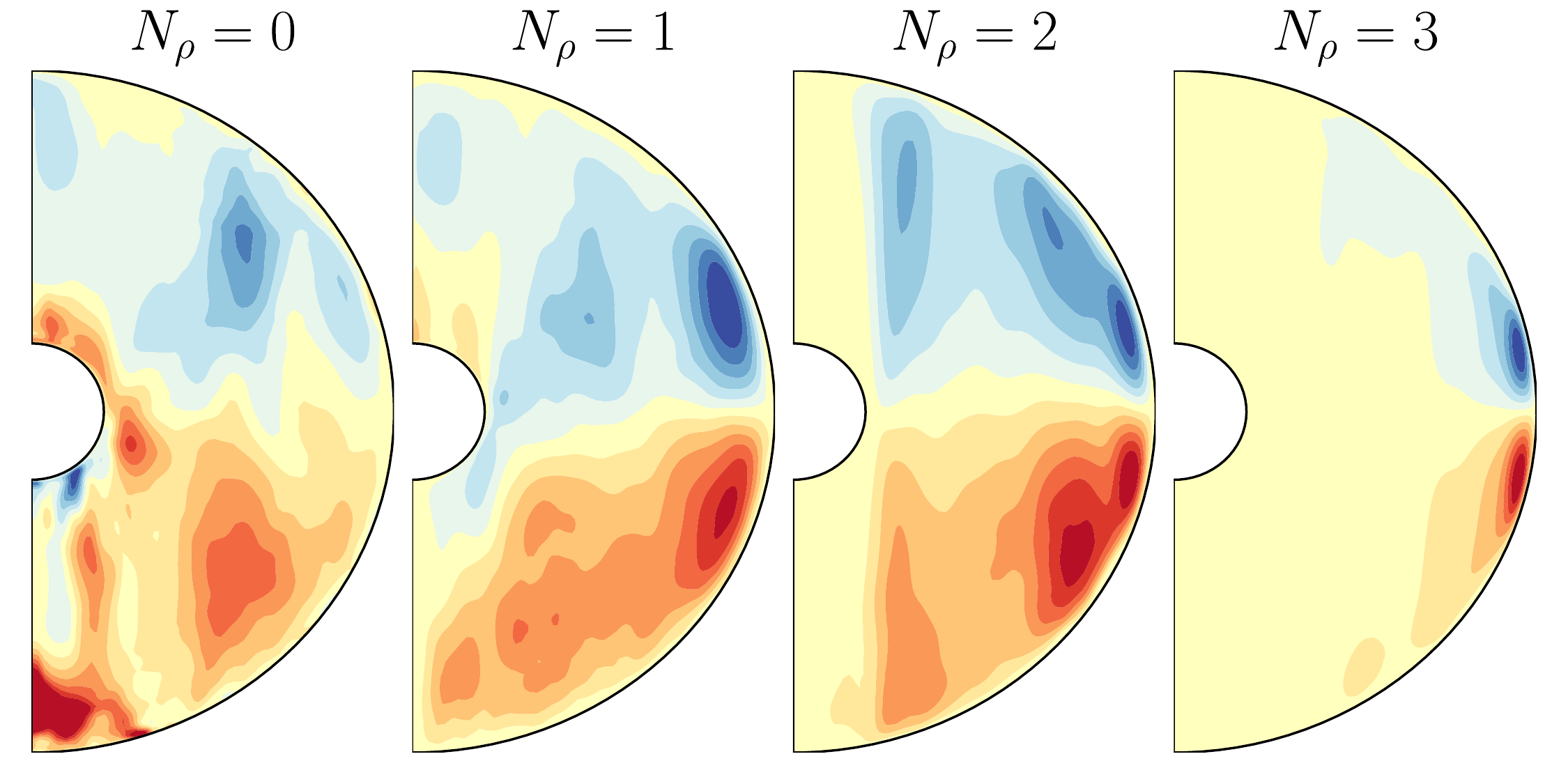}
 \includegraphics[width=9cm]{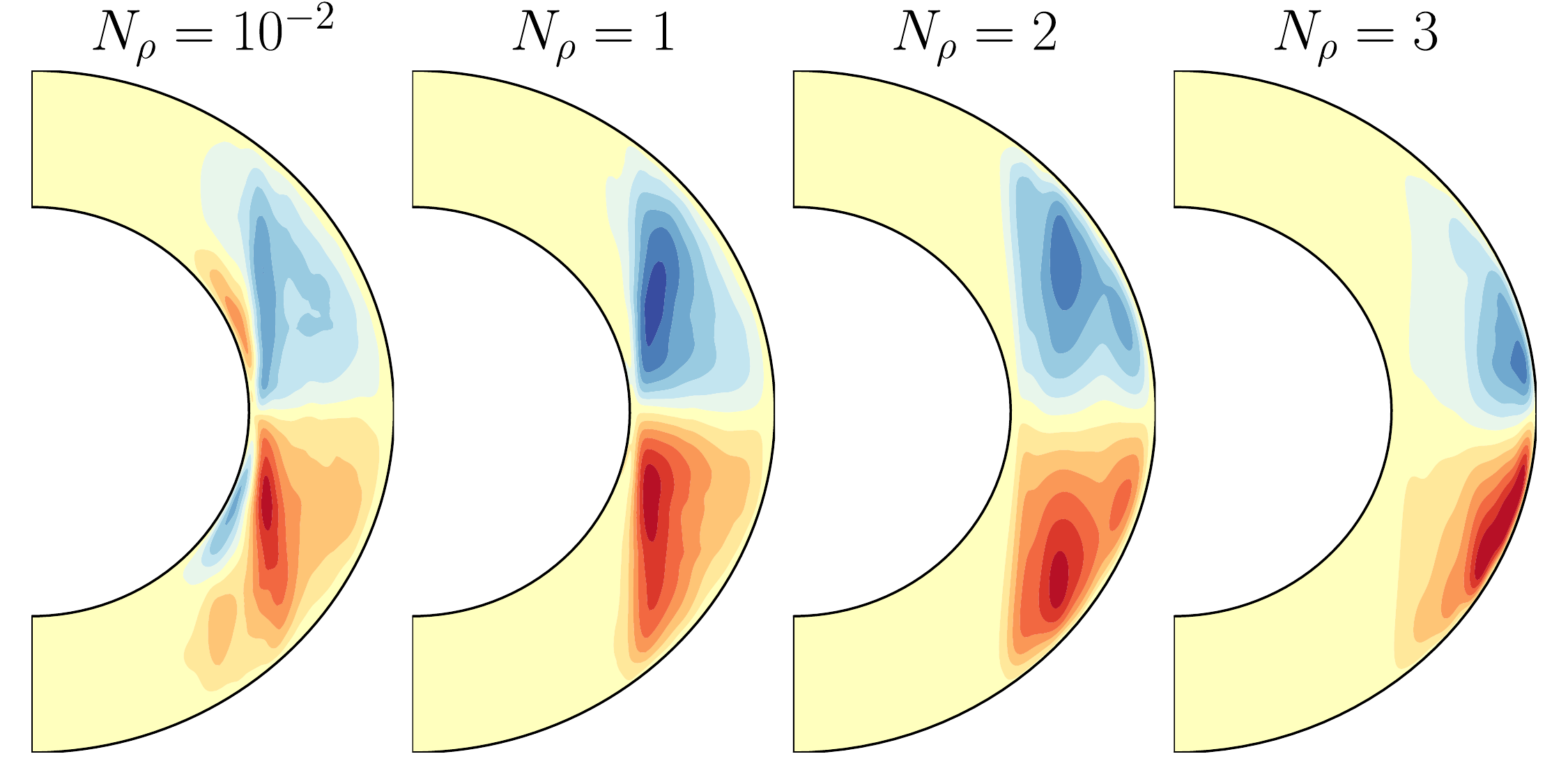}
 \caption{Azimuthal average of kinetic helicity
$\overline{\vec{u}'\cdot\vec{\nabla}\times\vec{u}'}$ for the models displayed
in Fig.~\ref{fig:vortz}. Positive (negative) values are rendered in red
(blue).}
 \label{fig:helicity}
\end{figure}

The kinetic helicity ${\cal H} = \overline{\vec{u'}\cdot\vec{\nabla}\times \vec{u'}}$
is a key ingredient in the induction process via the $\alpha$-effect
\citep[e.g.][]{Moffatt78}. Figure~\ref{fig:helicity} illustrates how
the helicity increasingly concentrates closer to the outer
boundary equator when the stratification intensifies simply because the
convective columns are the main carriers of helicity.
Mean field dynamo models have demonstrated that a similar concentration of the
$\alpha$-effect can promote non-axisymmetric dynamo modes of
low spherical harmonic order \citep[typically $m \sim
1-2$, see][]{Rudiger03,Jiang06}.
Non-axisymmetric $\alpha^2\Omega$ mean-field models also suggest a
lower critical dynamo number for non-axisymmetric dynamo modes
\citep[e.g.][]{Ruzmaikin88, Bassom05}.
These findings may explain the onset of the new larger wave number magnetic mode
in our strongly stratified cases.

\subsection{Parker waves}
\label{sec:parker}

\begin{figure}
 \centering
 \includegraphics[width=9cm]{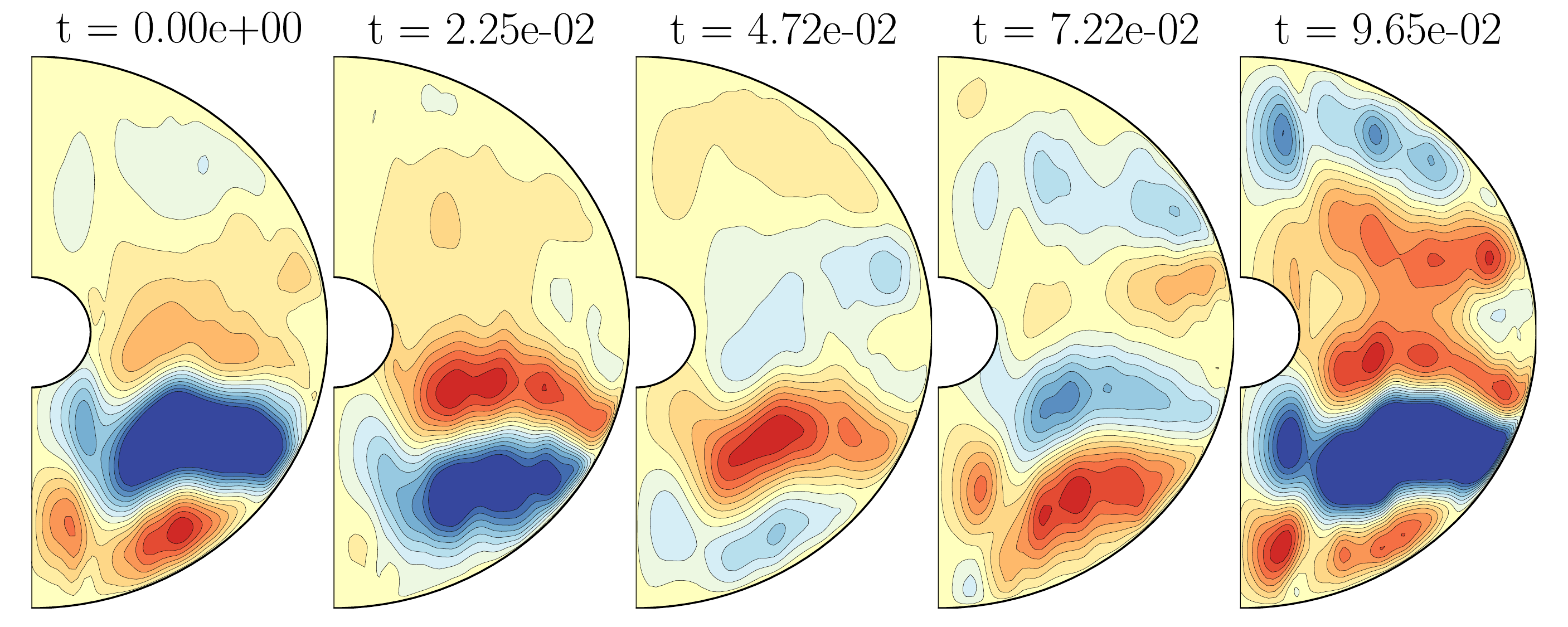}
 \includegraphics[width=9cm]{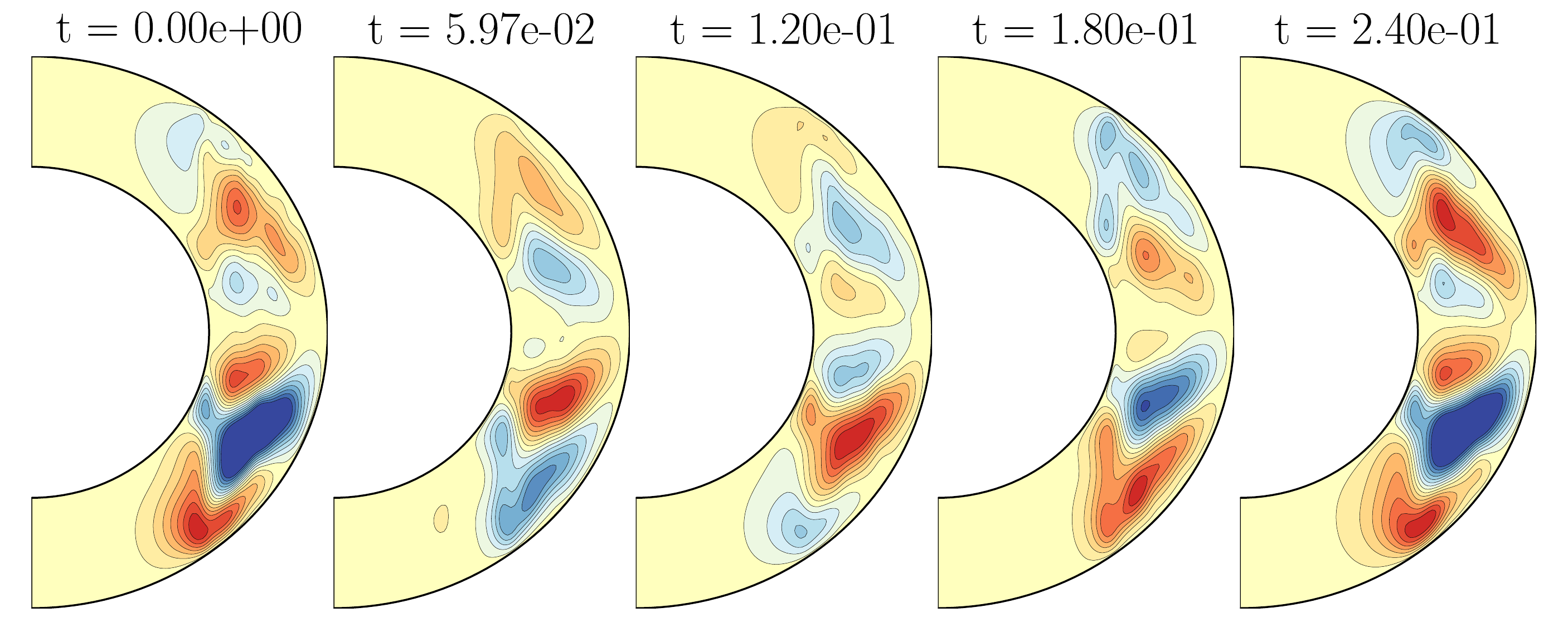}
 \caption{Time evolution of axisymmetric toroidal magnetic field
$\overline{B_\phi}$ for a simulation in a thick shell ($\eta=0.2$, $N_\rho=3$,
$\text{Ra}=1.1\times 10^{7}$, upper panel) and a simulation in a thin shell
($\eta=0.6$, $N_\rho=1$, $\text{Ra}=10^{6}$, lower panel). Times
are given here in magnetic diffusion units. Red (blue) correspond to positive
(negative) values.}
\label{fig:parker}
\end{figure}

\begin{figure}
 \centering
 \includegraphics[width=8cm]{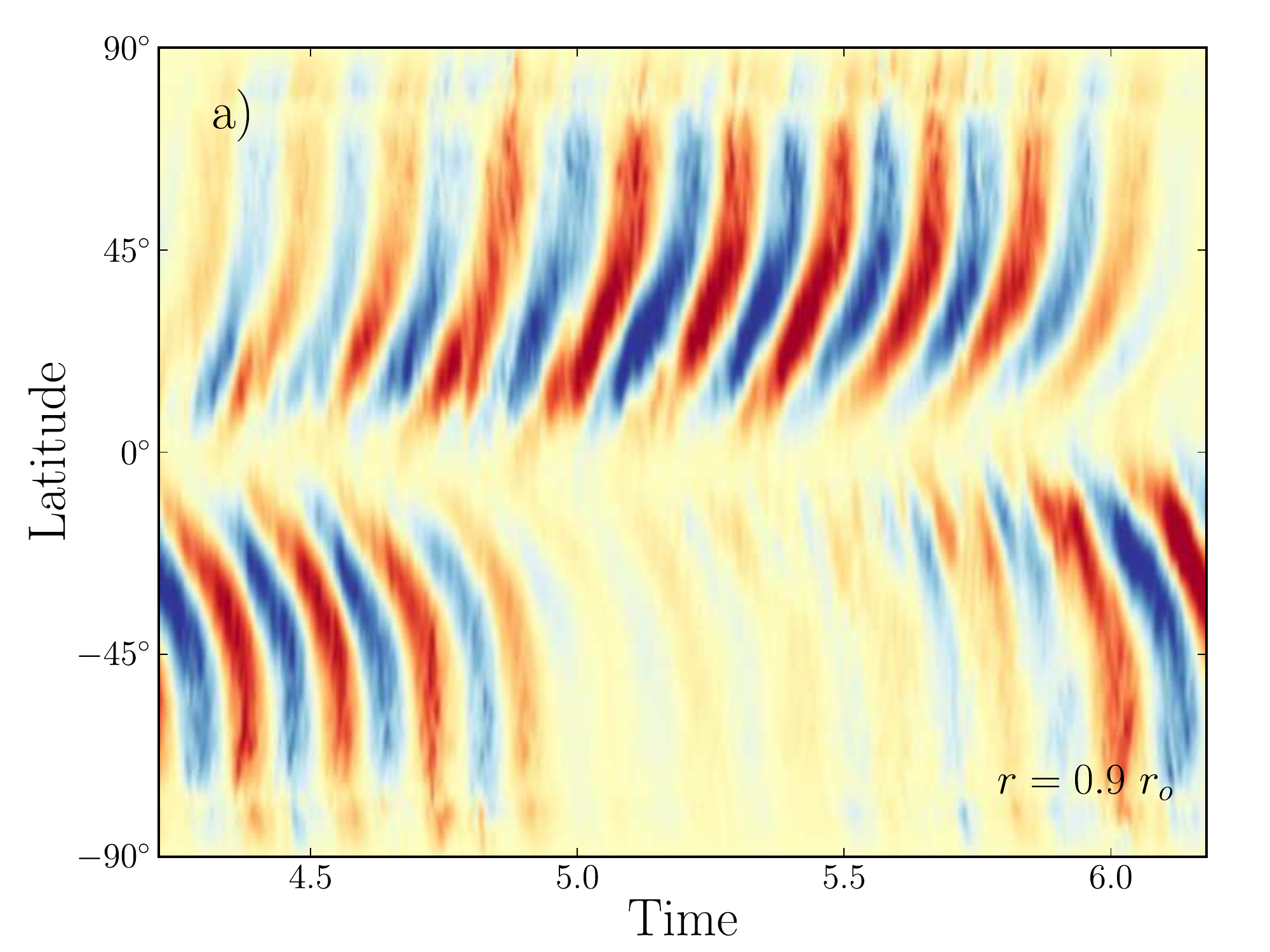}
 \includegraphics[width=8cm]{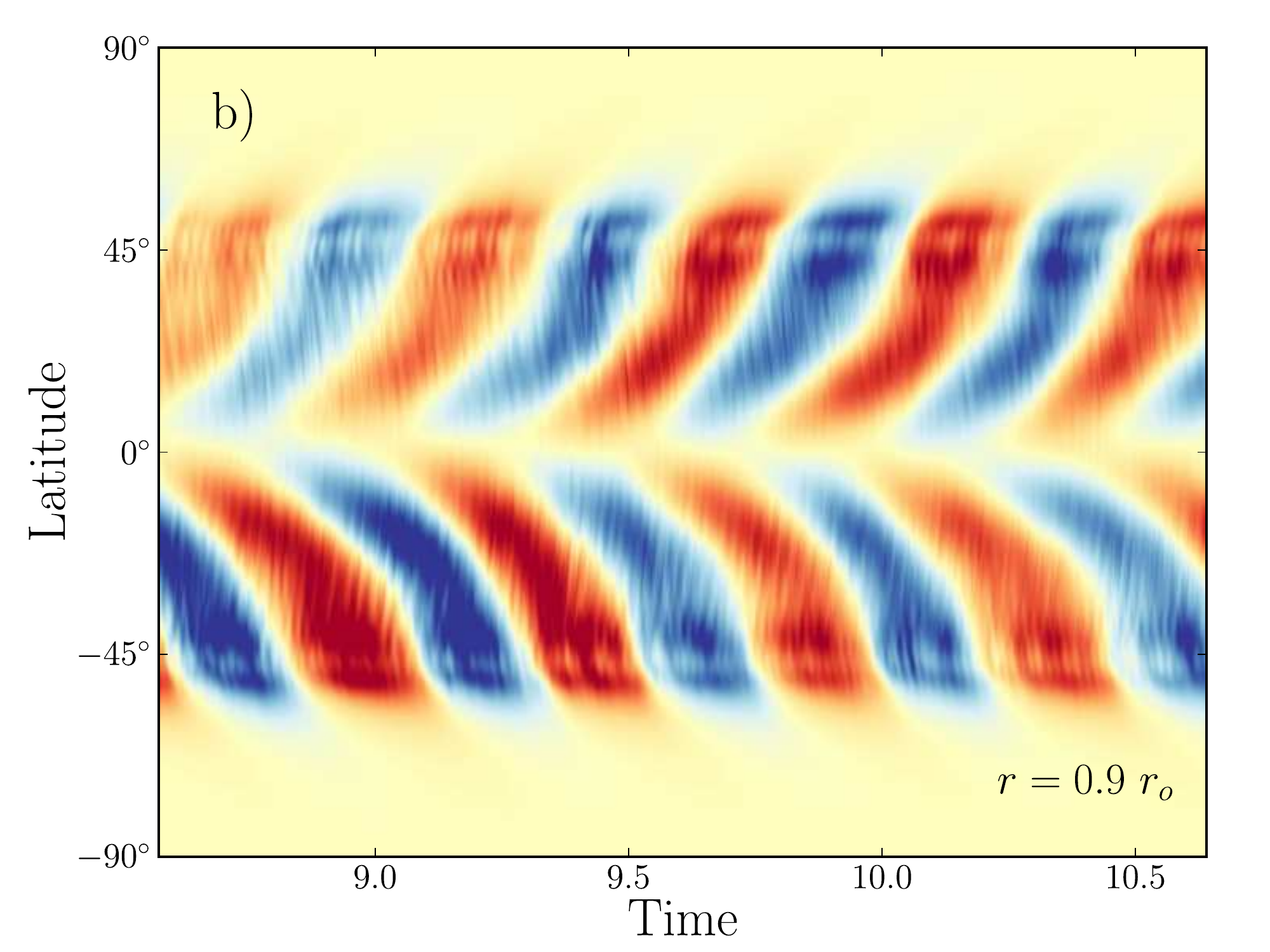}
 \caption{Butterfly diagrams: time evolution of axisymmetric toroidal
magnetic field $\overline{B_\phi}$ at $r=0.9 r_o$ for the two models of
Fig.~\ref{fig:parker}. Time is given here in magnetic diffusion units. Red 
(blue) correspond to positive (negative)
values.}
 \label{fig:butterfly}
\end{figure}

The  $\alpha\Omega$ or $\alpha^2\Omega$ dynamos on the multipolar
branch also show a distinct and interesting oscillatory
time dependence. Figures~\ref{fig:parker}-\ref{fig:butterfly} illustrate
two examples. The magnetic field is weak, multipolar and dominated
by non-axisymmetric components in both cases
(see for example Figs.~\ref{fig:brsurfEta02}-\ref{fig:brsurfEta06}).
For the thin shell (lower panels), a dipolar solution coexists
at identical parameters.
Toroidal magnetic field first emerges at low latitudes, then
travels towards the poles and finally vanishes roughly where the
tangent cylinder seems to prevent a further movement
towards even higher latitudes \citep{Schrinner11}.
When the patches have travelled half way, the cycle starts over with field of
opposite polarity appearing close to the equatorial plane.
The thick shell solution (top panels) demonstrates that during some
periods one hemisphere clearly dominates.
The characteristic period is roughly $0.1\tau_\lambda - 0.2\tau_\lambda$
where
$\tau_\lambda=d^2/\lambda$ is the magnetic diffusion time.
These are Parker dynamo waves, as we will demonstrate below,
very similar to those that have been previously identified in some Boussinesq
simulations \citep{Goudard08,Schrinner11,Simitev12}.

These coherent oscillations lead to the periodic patterns observed in the
``butterfly-diagram'' style illustrations shown in Fig.~\ref{fig:butterfly}.
The diagrams highlight the direction of travel which in Parker waves
is controlled by the gradient of zonal flows: they travel
polewards when like in our simulations the differential rotation decreases
with depth \citep[e.g.][]{Yoshimura75}.

The mean-field formalism allows to derive a dispersion relation for
these dynamo waves \citep{Busse06,Schrinner11}.
In this formalism, magnetic field and flow are
decomposed into the axisymmetric or mean contributions denoted by overbars

\begin{equation}
 \overline{\vec{B}} = \vec{B}_p+B\,\vec{e_\phi}=\vec{\nabla}\times (A\
\vec{e_\phi})+B\,\vec{e_\phi} \quad \text{;} \quad \overline{\vec{u}} = u\,
\vec{e_\phi},
\end{equation}
and the primed non-axisymmetric or fluctuating contributions
$\vec{B'}$ and $\vec{u'}$.
This allows to reformulate the axisymmetric part
of the induction equation (\ref{eq:induction}) to

\begin{equation}
\left\lbrace
\begin{aligned}
 \dfrac{\partial A}{\partial t}  & = \alpha B + \dfrac{1}{\text{Pm}}\nabla^2
A,
\\
 \dfrac{\partial B}{\partial t}  & = \vec{B}_p\cdot \vec{\nabla}\,u  +
(\vec{\nabla}\times \alpha\vec{B_p}) \cdot\vec{e_\phi} +
\dfrac{1}{\text{Pm}}\nabla^2 B,
\end{aligned}
\right.
\label{eq:meanfields}
\end{equation}
where the fluctuating mean electromotive force is  approximated
by an $\alpha$-effect, i.e.~by $\overline{\vec{u'}\times\vec{B'}}=\alpha\,
\vec{\overline{B}}$.

To simplify this system we assume a homogeneous $\alpha$ and also adopt the
plane layer formalism introduced by \cite{Parker55}, where the cartesian
coordinates $(x,y,z)$ correspond to $(\phi,\theta,r)$. Further
assuming that $\overline{u}$ depends only on $z$ (radius) leads to
\begin{equation}
\left\lbrace
\begin{aligned}
  \dfrac{\partial A}{\partial t}  & = \alpha B + \dfrac{1}{\text{Pm}}\nabla^2
A, \\
 \dfrac{\partial B}{\partial t}  & = -\alpha \nabla^2 A +
\dfrac{du}{dz} \dfrac{\partial A}{\partial y} + \dfrac{1}{\text{Pm}}\nabla^2
B.
\end{aligned}
\right.
\end{equation}
The ansatz $A, B \propto \exp(i(k_y y + k_z z)+\lambda t)$ with $\lambda = \tau
+i\omega$ allows to derive the following dispersion relation

\begin{equation}
 \left(\lambda + \dfrac{|\vec{k}|^2}{\text{Pm}}\right)^2 =\alpha^2\,|\vec{k}|^2
+ i k_y\,\alpha\,\dfrac{du}{dz}.
\end{equation}
For $\alpha\Omega$ dynamos we concentrated on in the following,
the imaginary part of $\lambda$ provides

\begin{equation}
 \omega = \lp \dfrac{1}{2}\,\alpha\,\dfrac{du}{dz}\,k_y \rp^{1/2}.
\end{equation}
To further simplify this dispersion relation we assume that
$z$ variations are of the order of the shell radius, i.e.~$k_y\sim 1/r_o$
and that the shear can be approximated by
$du/dz \sim \overline{u_\phi}/d \sim \text{Re}_{\text{zon}}$.

The $\alpha$ is particularly difficult to estimate.
A full derivation of the $\alpha$ tensor would, for example, require to
employ the test-field method \citep[e.g.][]{Schrinner07}.
For homogeneous and isotropic MHD turbulence, however, it may at least
crudely be approximated via the fluctuating kinetic helicity:

\begin{equation}
 \alpha = -\dfrac{\tau_c}{3}\,\overline{ \vec{u}'\cdot \nabla\times\vec{u}'},
 \label{eq:helicity}
\end{equation}
where $\tau_c$ is the typical lifetime of a convective feature
\citep[e.g.][]{Brandenburg05}. Following \cite{Brown10}, we use
$\tau_c=H_\rho/u'$ when the local density scale height
$H_\rho=(d\ln\rb/dr)^{-1}$ is smaller than $d$ and
$\tau_c=d/u'$ otherwise.
This finally leads to the simplified dispersion relation for Parker waves

\begin{equation}
 \omega \sim \lp \dfrac{\alpha\,\text{Re}_{\text{zon}}}{2\,r_o} \rp^{1/2}.
 \label{eq:freq}
\end{equation}

\begin{figure}
 \centering
 \includegraphics[width=9cm]{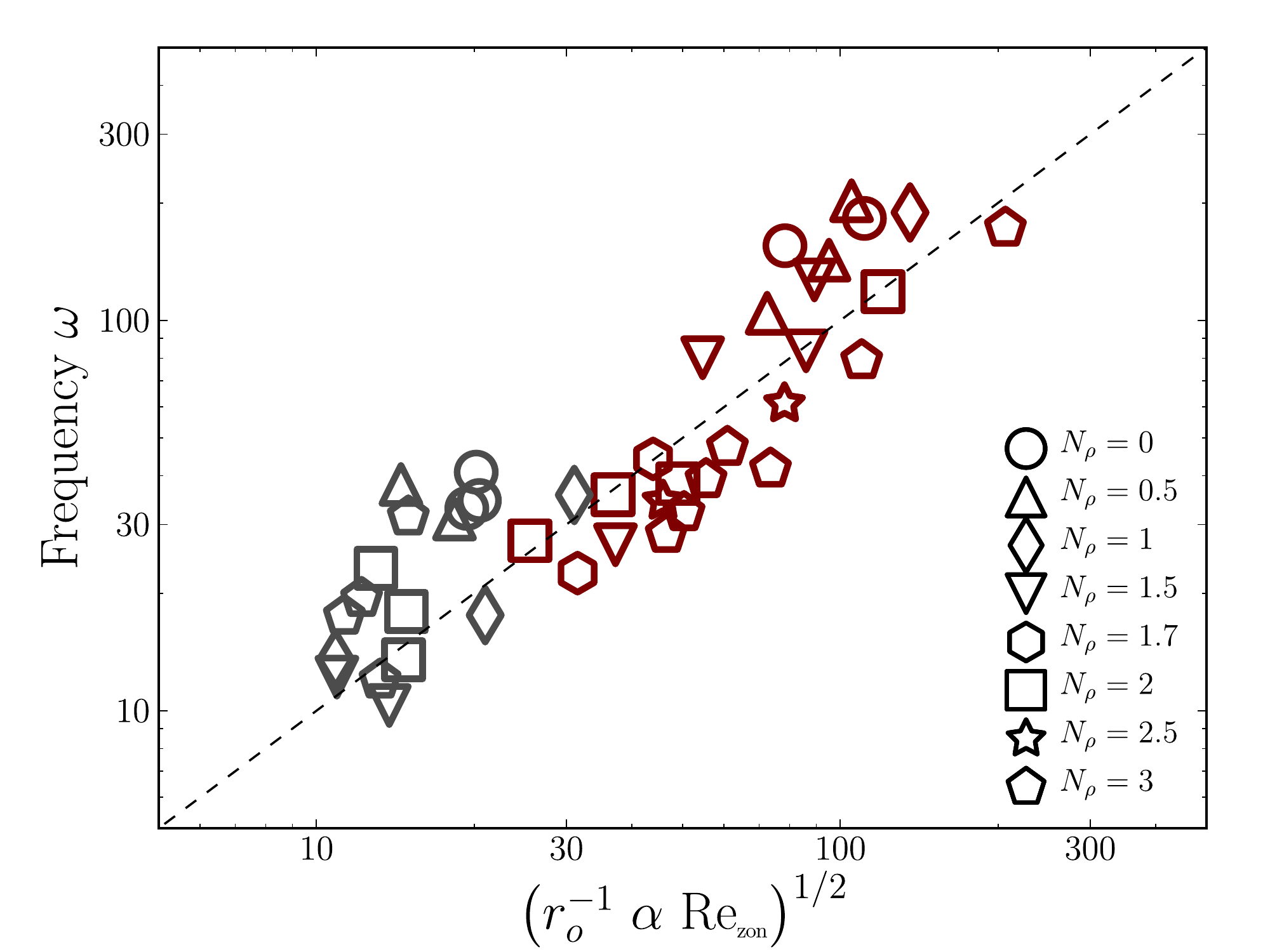}
 \caption{Frequencies of oscillatory dynamos plotted against the dispersion
relation of Parker waves given in Eq.~(\ref{eq:freq}). Red (grey) symbols
correspond to simulations in thick (thin) shells ($\eta=0.2$ and $\eta=0.6$,
respectively). Each type of symbol is associated to a given density
stratification.}
 \label{fig:freq}
\end{figure}

Figure~\ref{fig:freq} shows the comparison between the frequencies predicted
by this simplified dispersion relation and those found in
the numerical models. The later frequencies are obtained using a Fourier
transform of the butterfly diagrams displayed in Fig~\ref{fig:butterfly}. The
resulting $\widehat{B_\phi}(\theta, \omega)$ is then integrated over the
colatitude $\theta$ to derive the power spectrum. Even though individual
predictions fail by up to a factor two, the general agreement is quite
convincing
given the numerous approximation involved and is similar
to that reached in previous Boussinesq studies \citep{Busse06,Schrinner11}.
This confirms that the oscillations in our multipolar cases
are indeed Parker waves.
When the magnetic Reynolds number becomes too large ($\text{Rm} \gtrsim 150-200$),
the coherence of these oscillations is gradually lost and a definite frequency
is increasingly difficult to determine. This limits our analysis to
lower $\text{Re}_{\text{zon}}$ and $\alpha$ values and to frequencies
$\omega<300$.

\section{Discussion and conclusions}
\label{sec:discussion}

We have investigated the influence of background density stratification on
convection-driven dynamos in a rotating spherical shell.
The use of the anelastic approximation allowed us to exclude sound waves and 
the related short time steps \citep{Glatz1,Clune99,Jones09}.
Previous Boussinesq results have shown that inertial effects play
a decisive role for determining the magnetic field geometry.
When inertia becomes influential, only multipolar solutions with weak
magnetic fields are possible. When inertia is weak, two solutions can coexist:
dipole-dominated solutions with strong magnetic fields are found for
stress-free as well as rigid boundary conditions. For stress-free and
mixed boundary conditions, a multipolar branch is found at identical parameters
\citep[e.g.][]{Simitev09,Schrinner12}. Alternatively, the recent study by
\cite{King12} suggests that the transition between
dipolar and multipolar dynamos may occur when inertia becomes larger than
viscous forces. This transition is accompanied by an abrupt decrease of the
kinetic helicity. These results however seem  to be in
contradiction with previous studies where inertia is always larger than
viscosity \citep[e.g.][]{Wicht10} and further investigations are required to
clarify this contradiction.

Our anelastic simulations confirm this scenario for mild stratifications
corresponding to $N_\rho<1.8$. The reason for the bistability is a competition
between zonal winds and dipolar magnetic fields. Strong dipolar
magnetic fields  prevent significant zonal winds to develop. Strong
zonal winds, on the other hand prevent the production of significant
dipolar fields.
The two branches also differ in the induction mechanism. Strong zonal winds
promote an $\Omega$-effect which leads to a $\alpha\Omega$ or  $\alpha^2\Omega$
type of dynamo, while dipole-dominated magnetic fields are typically generated
in an $\alpha^2$ process.
The sizable axisymmetric toroidal fields produced by an $\Omega$-effect
typically lead to a coherent cyclic time evolution of the magnetic field
for moderate magnetic Reynolds numbers ($\text{Rm} < 200$).
This is consistent with previous Boussinesq studies \citep[e.g.][]{Schrinner07}
and numerical models of young solar-type stars \citep{Brown11} and has
been identified as Parker waves.
Contrary to what is observed in the solar cycle, these waves start at the
equator and travel towards the poles because of the opposite sign in the
zonal shear.

For stronger stratification with $N_\rho>1.8$,
the dipolar branch is lost and close to onset of dynamo action a
new magnetic mode characterised by a large wave number $m=2$ appears.
This is likely due to the concentration of the $\alpha$-effect
into a narrow region close the equator. According to mean-field
models, this would preferentially promote non-axisymmetric dynamos \citep[see
also][]{Chabrier06} with large wave numbers.
The collapse of the dipolar branch may explain the differences between
the weakly stratified and significantly dipolar simulations by \cite{Dobler06}
($\rho_{\text{bot}}/\rho_{\text{top}} \simeq 5$, i.e.~$N_\rho=1.6$) and
the strongly stratified anelastic and multipolar models by  \cite{Browning08}
($\rho_{\text{bot}}/\rho_{\text{top}} \simeq 100$, i.e.~$N_\rho=4.6$).

Numerical limitations force us to use excessively large diffusivities
in our simulations. Ekman numbers are thus orders of magnitude too large
and Reynolds number orders of magnitude too low.
When extrapolated to fast rotating planets and stars this type of models
nevertheless provides, for example, realistic magnetic field strengths
\citep{Olson06,Christensen09}. This gives us confidence to compare the observed
field geometries of planets and stars with predictions based on our simulation
results.

Our result are compatible with the dipole-dominated magnetic fields
on Jupiter and Saturn that are generated in their deeper metallic envelopes
where the density stratification is only mild \citep[roughly $N_\rho\sim
1.5-2$, see][]{Heimpel11,Nettelmann12}. Since the local Rossby number is very
small for these planets, however, bistability seems to be an option. This is
also the case for Uranus and Neptune, where $\text{Ro}_\ell$ and the density
contrast within the dynamo region are small \citep{Hubbard91,Olson06}. Their
multipolar magnetic field would then suggest that these planets would then
occupy the alternative branch offered by the bistability phenomenon.

Concerning rapidly rotating low-mass stars that may also fall into the low
$\text{Ro}_\ell$ regime, the spectropolarimetric observations of \cite{lateM}
suggest that late M stars with very similar parameters (mass and rotation rate)
come in two categories: some stars present a strong dipole-dominated magnetic
field while others show weaker and multipolar magnetic structures. These two
geometries may represent the two coexisting dynamo branches at smaller local
Rossby numbers.

For the bistability to be a viable explanation, however, our simulations
suggest that the dynamos must operate in a region with moderate
density stratification ($N_\rho \sim 1-2$) and supercriticality
($\text{Ra}/\text{Ra}_c \sim 10-30$). Since the stellar dynamos operate
presumably far from onset of convection, only multipolar fields would then be
possible. However, when decreasing the Ekman number towards more realistic
values, the simulations by \cite{Christensen06} suggest that the dipolar window
may persist at higher supercriticalities. In addition, since these stars have
huge density contrasts, our investigation would then generally predict
multipolar fields. A further exploration of the parameter space seems here
required to clarify this point. \cite{Stanley09}, for example, suggest that
lower Prandtl number may help in creating stronger dipole fields. Considering
radial-dependent properties (e.g. viscosity, thermal diffusivity and electrical
diffusivity) is also known to have a strong impact on the location of the
convective columns that could possibly help to avoid the concentration of
helicity close to the equator.

\begin{acknowledgements}
All the computations have been carried out on the GWDG computer facilities in
G\"ottingen. This work was supported by the Special Priority Program 1488
(PlanetMag, \url{http://www.planetmag.de}) of the German Science Foundation. It
is a pleasure to thank C.~A.~Jones for providing us his linear stability code.
\end{acknowledgements}

\bibliographystyle{aa}

{\small
\longtab{2}{%
\begin{longtable}{cccccccccccc}
\caption{Results table.}\\
\toprule
Model & $\eta$ & BC & $N_\rho$ & Rayleigh & $\text{Ra}/\text{Ra}_c$ & $f_{\text{dip}}$ & Rm & Ro & $\text{Ro}_\ell$ & $\Lambda$ & $\Lambda_\ell$ \\
\midrule
\endfirsthead
\caption{Continued.}\\
\toprule
Model & $\eta$ & BC & $N_\rho$ & Rayleigh & $\text{Ra}/\text{Ra}_c$ & $f_{\text{dip}}$ & Rm & Ro & $\text{Ro}_\ell$ & $\Lambda$ & $\Lambda_\ell$ \\
\midrule
\endhead
\bottomrule
\endfoot
1d & 0.2 & mixed &0.00 & $8.00 \times 10^{6}$ & 9.2 & 0.72& 122 & $6.57 \times 10^{-3}$ & $1.95 \times 10^{-2}$ & 1.3 & $7.13 \times 10^{-2}$ \\
2d & 0.2 & mixed &0.00 & $1.00 \times 10^{7}$ & 11.5 & 0.71& 143 & $7.77 \times 10^{-3}$ & $2.55 \times 10^{-2}$ & 2.2 & $1.14 \times 10^{-1}$ \\
2m & 0.2 & mixed &0.00 & $1.00 \times 10^{7}$ & 11.5 & 0.05& 180 & $9.55 \times 10^{-3}$ & $2.62 \times 10^{-2}$ & 0.6 & $3.33 \times 10^{-2}$ \\
3d & 0.2 & mixed &0.00 & $2.00 \times 10^{7}$ & 23.0 & 0.66& 252 & $1.36 \times 10^{-2}$ & $5.81 \times 10^{-2}$ & 5.6 & $2.24 \times 10^{-1}$ \\
3m & 0.2 & mixed & 0.00 & $2.00 \times 10^{7}$ & 23.0 & $1.62 \times 10^{-2}$ & 318 & $1.70 \times 10^{-2}$ & $6.26 \times 10^{-2}$ & 2.6 &  $9.73 \times 10^{-2}$\\
4d & 0.2 & mixed & 0.00 & $3.00 \times 10^{7}$ & 34.5 & $1.14 \times 10^{-2}$ & 424 & $2.26 \times 10^{-2}$ & $8.38 \times 10^{-2}$ & 4.5 &  $1.42 \times 10^{-1}$\\
5d & 0.2 & mixed & 0.00 & $4.00 \times 10^{7}$ & 45.9 & $1.05 \times 10^{-2}$ & 505 & $2.71 \times 10^{-2}$ & $1.05 \times 10^{-1}$ & 6.4 &  $1.78 \times 10^{-1}$\\
\midrule
6d & 0.2 & mixed &0.50 & $7.00 \times 10^{6}$ & 5.1 & 0.68& 104 & $5.26 \times 10^{-3}$ & $1.91 \times 10^{-2}$ & 1.6 & $8.68 \times 10^{-2}$ \\
7d & 0.2 & mixed &0.50 & $1.00 \times 10^{7}$ & 7.3 & 0.71& 146 & $7.39 \times 10^{-3}$ & $3.28 \times 10^{-2}$ & 4.0 & $1.84 \times 10^{-1}$ \\
7m & 0.2 & mixed &0.50 & $1.00 \times 10^{7}$ & 7.3 & 0.02& 185 & $9.35 \times 10^{-3}$ & $3.75 \times 10^{-2}$ & 1.0 & $4.93 \times 10^{-2}$ \\
8d & 0.2 & mixed &0.50 & $1.50 \times 10^{7}$ & 10.9 & 0.67& 220 & $1.10 \times 10^{-2}$ & $5.52 \times 10^{-2}$ & 7.3 & $2.83 \times 10^{-1}$ \\
8m & 0.2 & mixed &0.50 & $1.50 \times 10^{7}$ & 10.9 & 0.03& 285 & $1.44 \times 10^{-2}$ & $6.24 \times 10^{-2}$ & 2.5 & $9.41 \times 10^{-2}$ \\
9d & 0.2 & mixed &0.50 & $1.90 \times 10^{7}$ & 13.8 & 0.02& 351 & $1.77 \times 10^{-2}$ & $7.77 \times 10^{-2}$ & 3.7 & $1.18 \times 10^{-1}$ \\
10d & 0.2 & mixed & 0.50 & $2.00 \times 10^{7}$ & 14.5 & $1.82 \times 10^{-2}$ & 364 & $1.83 \times 10^{-2}$ & $8.16 \times 10^{-2}$ & 4.0 &  $1.26 \times 10^{-1}$\\
11d & 0.2 & mixed & 0.50 & $3.00 \times 10^{7}$ & 21.8 & $1.05 \times 10^{-2}$ & 491 & $2.47 \times 10^{-2}$ & $1.10 \times 10^{-1}$ & 7.0 &  $1.78 \times 10^{-1}$\\
\midrule
12d & 0.2 & mixed &1.00 & $6.00 \times 10^{6}$ & 3.1 & 0.86& 71 & $3.59 \times 10^{-3}$ & $1.55 \times 10^{-2}$ & 0.8 & $4.76 \times 10^{-2}$ \\
13d & 0.2 & mixed &1.00 & $8.00 \times 10^{6}$ & 4.1 & 0.67& 117 & $5.87 \times 10^{-3}$ & $3.03 \times 10^{-2}$ & 3.4 & $1.62 \times 10^{-1}$ \\
13m & 0.2 & mixed &1.00 & $8.00 \times 10^{6}$ & 4.1 & 0.70& 115 & $5.81 \times 10^{-3}$ & $2.96 \times 10^{-2}$ & 3.2 & $1.43 \times 10^{-1}$ \\
14d & 0.2 & mixed &1.00 & $1.00 \times 10^{7}$ & 5.2 & 0.68& 152 & $7.65 \times 10^{-3}$ & $4.18 \times 10^{-2}$ & 5.7 & $2.37 \times 10^{-1}$ \\
14m & 0.2 & mixed &1.00 & $1.00 \times 10^{7}$ & 5.2 & 0.69& 149 & $7.49 \times 10^{-3}$ & $4.04 \times 10^{-2}$ & 6.2 & $2.39 \times 10^{-1}$ \\
15d & 0.2 & mixed &1.00 & $1.50 \times 10^{7}$ & 7.8 & 0.52& 256 & $1.29 \times 10^{-2}$ & $7.44 \times 10^{-2}$ & 9.0 & $2.83 \times 10^{-1}$ \\
15m & 0.2 & mixed &1.00 & $1.50 \times 10^{7}$ & 7.8 & 0.03& 299 & $1.50 \times 10^{-2}$ & $7.51 \times 10^{-2}$ & 3.8 & $1.13 \times 10^{-1}$ \\
16d & 0.2 & mixed & 1.00 & $2.00 \times 10^{7}$ & 10.3 & $1.40 \times 10^{-2}$ & 390 & $1.96 \times 10^{-2}$ & $1.01 \times 10^{-1}$ & 6.3 &  $1.61 \times 10^{-1}$\\
\midrule
17d & 0.2 & mixed &1.50 & $7.00 \times 10^{6}$ & 2.7 & 0.81& 85 & $4.29 \times 10^{-3}$ & $2.61 \times 10^{-2}$ & 2.0 & $9.12 \times 10^{-2}$ \\
17m & 0.2 & mixed &1.50 & $7.00 \times 10^{6}$ & 2.7 & 0.15& 88 & $4.56 \times 10^{-3}$ & $2.05 \times 10^{-2}$ & 0.5 & $2.26 \times 10^{-2}$ \\
18d & 0.2 & mixed &1.50 & $8.00 \times 10^{6}$ & 3.1 & 0.75& 112 & $5.63 \times 10^{-3}$ & $3.57 \times 10^{-2}$ & 3.5 & $1.53 \times 10^{-1}$ \\
18m & 0.2 & mixed &1.50 & $8.00 \times 10^{6}$ & 3.1 & 0.13& 126 & $6.36 \times 10^{-3}$ & $3.59 \times 10^{-2}$ & 1.4 & $5.13 \times 10^{-2}$ \\
19d & 0.2 & mixed &1.50 & $1.00 \times 10^{7}$ & 3.8 & 0.63& 165 & $8.29 \times 10^{-3}$ & $5.28 \times 10^{-2}$ & 5.8 & $2.03 \times 10^{-1}$ \\
19m & 0.2 & mixed &1.50 & $1.00 \times 10^{7}$ & 3.8 & 0.05& 184 & $9.27 \times 10^{-3}$ & $5.32 \times 10^{-2}$ & 2.4 & $7.63 \times 10^{-2}$ \\
20d & 0.2 & mixed &1.50 & $1.30 \times 10^{7}$ & 5.0 & 0.05& 256 & $1.29 \times 10^{-2}$ & $7.05 \times 10^{-2}$ & 3.7 & $9.77 \times 10^{-2}$ \\
21d & 0.2 & mixed & 1.50 & $1.40 \times 10^{7}$ & 5.4 & $1.72 \times 10^{-2}$ & 278 & $1.40 \times 10^{-2}$ & $7.50 \times 10^{-2}$ & 3.9 &  $9.63 \times 10^{-2}$\\
22d & 0.2 & mixed & 1.50 & $1.50 \times 10^{7}$ & 5.7 & $1.57 \times 10^{-2}$ & 299 & $1.50 \times 10^{-2}$ & $8.18 \times 10^{-2}$ & 4.5 &  $1.07 \times 10^{-1}$\\
23d & 0.2 & mixed & 1.50 & $2.00 \times 10^{7}$ & 7.6 & $1.82 \times 10^{-2}$ & 410 & $2.06 \times 10^{-2}$ & $1.17 \times 10^{-1}$ & 8.5 &  $1.76 \times 10^{-1}$\\
\midrule
24d & 0.2 & mixed &1.70 & $9.00 \times 10^{6}$ & 3.1 & 0.73& 134 & $6.73 \times 10^{-3}$ & $4.51 \times 10^{-2}$ & 4.4 & $1.69 \times 10^{-1}$ \\
24m & 0.2 & mixed &1.70 & $9.00 \times 10^{6}$ & 3.1 & 0.07& 153 & $7.71 \times 10^{-3}$ & $4.62 \times 10^{-2}$ & 2.0 & $7.00 \times 10^{-2}$ \\
25d & 0.2 & mixed &1.70 & $1.20 \times 10^{7}$ & 4.1 & 0.06& 230 & $1.16 \times 10^{-2}$ & $6.58 \times 10^{-2}$ & 3.4 & $8.84 \times 10^{-2}$ \\
\midrule
26d & 0.2 & mixed & 2.00 & $7.50 \times 10^{6}$ & 2.2 & $1.97 \times 10^{-2}$ & 87 & $4.42 \times 10^{-3}$ & $2.31 \times 10^{-2}$ & 0.5 &  $2.27 \times 10^{-2}$\\
27d & 0.2 & mixed & 2.00 & $8.70 \times 10^{6}$ & 2.5 & $1.22 \times 10^{-2}$ & 128 & $6.48 \times 10^{-3}$ & $3.80 \times 10^{-2}$ & 1.4 &  $4.76 \times 10^{-2}$\\
28d & 0.2 & mixed & 2.00 & $1.00 \times 10^{7}$ & 2.9 & $1.72 \times 10^{-2}$ & 165 & $8.34 \times 10^{-3}$ & $4.91 \times 10^{-2}$ & 2.1 &  $6.48 \times 10^{-2}$\\
29d & 0.2 & mixed & 2.00 & $1.50 \times 10^{7}$ & 4.3 & $1.37 \times 10^{-2}$ & 294 & $1.48 \times 10^{-2}$ & $8.89 \times 10^{-2}$ & 5.3 &  $1.16 \times 10^{-1}$\\
30d & 0.2 & mixed & 2.00 & $2.00 \times 10^{7}$ & 5.8 & $1.82 \times 10^{-2}$ & 415 & $2.08 \times 10^{-2}$ & $1.37 \times 10^{-1}$ & 12.8 &  $2.01 \times 10^{-1}$\\
\midrule
31d & 0.2 & mixed & 2.50 & $1.00 \times 10^{7}$ & 2.2 & $4.74 \times 10^{-3}$ & 150 & $7.61 \times 10^{-3}$ & $4.54 \times 10^{-2}$ & 2.1 &  $5.63 \times 10^{-2}$\\
32d & 0.2 & mixed & 2.50 & $1.50 \times 10^{7}$ & 3.3 & $8.63 \times 10^{-3}$ & 282 & $1.42 \times 10^{-2}$ & $9.26 \times 10^{-2}$ & 6.3 &  $1.21 \times 10^{-1}$\\
\midrule
33d & 0.2 & mixed & 3.00 & $1.05 \times 10^{7}$ & 2.3 & $2.23 \times 10^{-3}$ & 144 & $7.28 \times 10^{-3}$ & $3.75 \times 10^{-2}$ & 1.3 &  $3.21 \times 10^{-2}$\\
34d & 0.2 & mixed & 3.00 & $1.10 \times 10^{7}$ & 2.4 & $3.60 \times 10^{-3}$ & 157 & $7.92 \times 10^{-3}$ & $4.87 \times 10^{-2}$ & 2.4 &  $5.66 \times 10^{-2}$\\
35d & 0.2 & mixed & 3.00 & $1.20 \times 10^{7}$ & 2.6 & $5.60 \times 10^{-3}$ & 184 & $9.30 \times 10^{-3}$ & $6.13 \times 10^{-2}$ & 4.0 &  $8.27 \times 10^{-2}$\\
36d & 0.2 & mixed & 3.00 & $1.30 \times 10^{7}$ & 2.8 & $9.43 \times 10^{-3}$ & 211 & $1.06 \times 10^{-2}$ & $7.61 \times 10^{-2}$ & 5.8 &  $1.05 \times 10^{-1}$\\
37d & 0.2 & mixed & 3.00 & $1.50 \times 10^{7}$ & 3.2 & $1.84 \times 10^{-2}$ & 261 & $1.31 \times 10^{-2}$ & $9.94 \times 10^{-2}$ & 8.5 &  $1.46 \times 10^{-1}$\\
38d & 0.2 & mixed & 3.00 & $2.00 \times 10^{7}$ & 4.3 & $8.90 \times 10^{-3}$ & 372 & $1.86 \times 10^{-2}$ & $1.46 \times 10^{-1}$ & 15.5 &  $1.68 \times 10^{-1}$\\
39d & 0.2 & mixed & 3.00 & $4.00 \times 10^{7}$ & 8.6 & $1.34 \times 10^{-2}$ & 694 & $3.47 \times 10^{-2}$ & $2.67 \times 10^{-1}$ & 47.3 &  $3.18 \times 10^{-1}$\\
\midrule
\midrule
40d & 0.6 & stress-free &0.01 & $6.00 \times 10^{5}$ & 3.5 & 0.91& 45 & $2.27 
\times 10^{-3}$ & $1.43 \times 10^{-2}$ & 4.0 & $1.21$ \\
40m & 0.6 & stress-free & 0.01 & $6.00 \times 10^{5}$ & 3.5 & $1.06 \times 10^{-4}$ & 64 & $3.28 \times 10^{-3}$ & $9.78 \times 10^{-3}$ & 0.0 &  $2.86 \times 10^{-3}$\\
41d & 0.6 & stress-free &0.01 & $9.00 \times 10^{5}$ & 5.2 & 0.85& 71 & $3.56 \times 10^{-3}$ & $3.15 \times 10^{-2}$ & 0.3 & $8.39 \times 10^{-2}$ \\
41m & 0.6 & stress-free & 0.01 & $9.00 \times 10^{5}$ & 5.2 & $2.63 \times 10^{-3}$ & 74 & $3.74 \times 10^{-3}$ & $2.81 \times 10^{-2}$ & 0.2 &  $2.72 \times 10^{-2}$\\
42d & 0.6 & stress-free &0.01 & $1.00 \times 10^{6}$ & 5.8 & 0.81& 80 & $4.01 \times 10^{-3}$ & $3.68 \times 10^{-2}$ & 0.4 & $9.89 \times 10^{-2}$ \\
42m & 0.6 & stress-free & 0.01 & $1.00 \times 10^{6}$ & 5.8 & $2.16 \times 10^{-3}$ & 84 & $4.24 \times 10^{-3}$ & $3.34 \times 10^{-2}$ & 0.2 &  $3.09 \times 10^{-2}$\\
43d & 0.6 & stress-free &0.01 & $2.00 \times 10^{6}$ & 11.5 & 0.75& 157 & $7.86 \times 10^{-3}$ & $7.02 \times 10^{-2}$ & 3.8 & $4.82 \times 10^{-1}$ \\
43m & 0.6 & stress-free & 0.01 & $2.00 \times 10^{6}$ & 11.5 & $7.46 \times 10^{-3}$ & 203 & $1.02 \times 10^{-2}$ & $8.98 \times 10^{-2}$ & 1.0 &  $1.17 \times 10^{-1}$\\
44d & 0.6 & stress-free &0.01 & $3.00 \times 10^{6}$ & 17.3 & 0.64& 241 & $1.21 \times 10^{-2}$ & $1.13 \times 10^{-1}$ & 8.5 & $7.95 \times 10^{-1}$ \\
44m & 0.6 & stress-free & 0.01 & $3.00 \times 10^{6}$ & 17.3 & $4.32 \times 10^{-3}$ & 317 & $1.58 \times 10^{-2}$ & $1.39 \times 10^{-1}$ & 3.0 &  $2.56 \times 10^{-1}$\\
45d & 0.6 & stress-free & 0.01 & $4.00 \times 10^{6}$ & 23.0 & $2.20 \times 10^{-3}$ & 403 & $2.02 \times 10^{-2}$ & $1.76 \times 10^{-1}$ & 5.1 &  $3.67 \times 10^{-1}$\\
\midrule
46d & 0.6 & stress-free &0.50 & $8.00 \times 10^{5}$ & 2.6 & 0.82& 58 & $2.96 \times 10^{-3}$ & $2.60 \times 10^{-2}$ & 4.7 & $9.24 \times 10^{-1}$ \\
46m & 0.6 & stress-free & 0.50 & $8.00 \times 10^{5}$ & 2.6 & $7.09 \times 10^{-3}$ & 50 & $2.53 \times 10^{-3}$ & $2.60 \times 10^{-2}$ & 0.1 &  $2.31 \times 10^{-2}$\\
47d & 0.6 & stress-free &0.50 & $1.00 \times 10^{6}$ & 3.2 & 0.80& 73 & $3.71 \times 10^{-3}$ & $3.20 \times 10^{-2}$ & 5.4 & $8.90 \times 10^{-1}$ \\
47m & 0.6 & stress-free & 0.50 & $1.00 \times 10^{6}$ & 3.2 & $2.59 \times 10^{-3}$ & 69 & $3.47 \times 10^{-3}$ & $3.79 \times 10^{-2}$ & 0.2 &  $3.57 \times 10^{-2}$\\
48d & 0.6 & stress-free &0.50 & $2.00 \times 10^{6}$ & 6.4 & 0.64& 164 & $8.23 \times 10^{-3}$ & $7.99 \times 10^{-2}$ & 8.3 & $7.80 \times 10^{-1}$ \\
48m & 0.6 & stress-free & 0.50 & $2.00 \times 10^{6}$ & 6.4 & $1.03 \times 10^{-2}$ & 195 & $9.77 \times 10^{-3}$ & $1.01 \times 10^{-1}$ & 1.2 &  $1.25 \times 10^{-1}$\\
49d & 0.6 & stress-free &0.50 & $3.00 \times 10^{6}$ & 9.6 & 0.53& 276 & $1.39 \times 10^{-2}$ & $1.43 \times 10^{-1}$ & 8.8 & $6.52 \times 10^{-1}$ \\
49m & 0.6 & stress-free & 0.50 & $3.00 \times 10^{6}$ & 9.6 & $1.10 \times 10^{-3}$ & 313 & $1.58 \times 10^{-2}$ & $1.53 \times 10^{-1}$ & 3.7 &  $2.87 \times 10^{-1}$\\
50d & 0.6 & stress-free & 0.50 & $4.00 \times 10^{6}$ & 12.8 & $1.86 \times 10^{-3}$ & 414 & $2.08 \times 10^{-2}$ & $2.00 \times 10^{-1}$ & 6.5 &  $4.05 \times 10^{-1}$\\
51d & 0.6 & stress-free & 0.50 & $5.00 \times 10^{6}$ & 16.0 & $9.86 \times 10^{-4}$ & 507 & $2.55 \times 10^{-2}$ & $2.45 \times 10^{-1}$ & 9.5 &  $5.05 \times 10^{-1}$\\
\midrule
52d & 0.6 & stress-free &1.00 & $1.00 \times 10^{6}$ & 1.9 & 0.73& 64 & $3.23 \times 10^{-3}$ & $3.68 \times 10^{-2}$ & 3.9 & $5.72 \times 10^{-1}$ \\
52m & 0.6 & stress-free & 1.00 & $1.00 \times 10^{6}$ & 1.9 & $1.50 \times 10^{-2}$ & 50 & $2.51 \times 10^{-3}$ & $3.49 \times 10^{-2}$ & 0.2 &  $2.21 \times 10^{-2}$\\
53d & 0.6 & stress-free &1.00 & $1.50 \times 10^{6}$ & 2.9 & 0.68& 113 & $5.70 \times 10^{-3}$ & $6.98 \times 10^{-2}$ & 3.6 & $4.54 \times 10^{-1}$ \\
53m & 0.6 & stress-free & 1.00 & $1.50 \times 10^{6}$ & 2.9 & $1.71 \times 10^{-2}$ & 109 & $5.47 \times 10^{-3}$ & $6.81 \times 10^{-2}$ & 0.6 &  $6.88 \times 10^{-2}$\\
54d & 0.6 & stress-free &1.00 & $2.00 \times 10^{6}$ & 3.9 & 0.62& 149 & $7.50 \times 10^{-3}$ & $8.02 \times 10^{-2}$ & 9.4 & $8.20 \times 10^{-1}$ \\
54m & 0.6 & stress-free & 1.00 & $2.00 \times 10^{6}$ & 3.9 & $1.55 \times 10^{-2}$ & 173 & $8.67 \times 10^{-3}$ & $1.02 \times 10^{-1}$ & 1.4 &  $1.29 \times 10^{-1}$\\
55d & 0.6 & stress-free &1.00 & $3.00 \times 10^{6}$ & 5.8 & 0.49& 275 & $1.38 \times 10^{-2}$ & $1.59 \times 10^{-1}$ & 8.4 & $5.65 \times 10^{-1}$ \\
55m & 0.6 & stress-free & 1.00 & $3.00 \times 10^{6}$ & 5.8 & $3.45 \times 10^{-3}$ & 287 & $1.44 \times 10^{-2}$ & $1.59 \times 10^{-1}$ & 4.0 &  $2.71 \times 10^{-1}$\\
56d & 0.6 & stress-free & 1.00 & $4.00 \times 10^{6}$ & 7.7 & $4.90 \times 10^{-3}$ & 378 & $1.89 \times 10^{-2}$ & $2.03 \times 10^{-1}$ & 6.9 &  $3.94 \times 10^{-1}$\\
57d & 0.6 & stress-free & 1.00 & $5.00 \times 10^{6}$ & 9.7 & $1.37 \times 10^{-3}$ & 465 & $2.33 \times 10^{-2}$ & $2.47 \times 10^{-1}$ & 10.1 &  $5.03 \times 10^{-1}$\\
\midrule
58d & 0.6 & stress-free &1.50 & $1.20 \times 10^{6}$ & 1.5 & 0.66& 58 & $2.93 \times 10^{-3}$ & $3.56 \times 10^{-2}$ & 2.9 & $3.79 \times 10^{-1}$ \\
58m & 0.6 & stress-free &1.50 & $1.20 \times 10^{6}$ & 1.5 & 0.06& 46 & $2.32 \times 10^{-3}$ & $3.03 \times 10^{-2}$ & 0.2 & $3.03 \times 10^{-2}$ \\
59d & 0.6 & stress-free &1.50 & $1.50 \times 10^{6}$ & 1.8 & 0.66& 89 & $4.47 \times 10^{-3}$ & $6.08 \times 10^{-2}$ & 3.5 & $3.77 \times 10^{-1}$ \\
59m & 0.6 & stress-free &1.50 & $1.50 \times 10^{6}$ & 1.8 & 0.07& 81 & $4.09 \times 10^{-3}$ & $5.64 \times 10^{-2}$ & 0.5 & $4.87 \times 10^{-2}$ \\
60d & 0.6 & stress-free &1.50 & $2.00 \times 10^{6}$ & 2.5 & 0.49& 146 & $7.33 \times 10^{-3}$ & $1.02 \times 10^{-1}$ & 3.2 & $2.54 \times 10^{-1}$ \\
60m & 0.6 & stress-free &1.50 & $2.00 \times 10^{6}$ & 2.5 & 0.04& 142 & $7.12 \times 10^{-3}$ & $9.28 \times 10^{-2}$ & 1.3 & $1.05 \times 10^{-1}$ \\
61d & 0.6 & stress-free & 1.50 & $2.50 \times 10^{6}$ & 3.1 & $1.45 \times 10^{-2}$ & 202 & $1.01 \times 10^{-2}$ & $1.25 \times 10^{-1}$ & 3.2 &  $2.27 \times 10^{-1}$\\
62d & 0.6 & stress-free & 1.50 & $3.00 \times 10^{6}$ & 3.7 & $1.46 \times 10^{-2}$ & 253 & $1.26 \times 10^{-2}$ & $1.52 \times 10^{-1}$ & 4.8 &  $2.93 \times 10^{-1}$\\
63d & 0.6 & stress-free & 1.50 & $4.00 \times 10^{6}$ & 4.9 & $7.82 \times 10^{-3}$ & 340 & $1.70 \times 10^{-2}$ & $1.97 \times 10^{-1}$ & 8.2 &  $4.10 \times 10^{-1}$\\
\midrule
64d & 0.6 & stress-free & 2.00 & $1.80 \times 10^{6}$ & 1.6 & $1.73 \times 10^{-2}$ & 83 & $4.16 \times 10^{-3}$ & $6.31 \times 10^{-2}$ & 0.8 &  $7.28 \times 10^{-2}$\\
65d & 0.6 & stress-free &2.00 & $2.00 \times 10^{6}$ & 1.8 & 0.03& 105 & $5.26 \times 10^{-3}$ & $7.84 \times 10^{-2}$ & 1.0 & $7.74 \times 10^{-2}$ \\
66d & 0.6 & stress-free & 2.00 & $2.20 \times 10^{6}$ & 1.9 & $1.00 \times 10^{-2}$ & 126 & $6.35 \times 10^{-3}$ & $9.50 \times 10^{-2}$ & 1.5 &  $1.06 \times 10^{-1}$\\
67d & 0.6 & stress-free &2.00 & $2.50 \times 10^{6}$ & 2.2 & 0.02& 156 & $7.81 \times 10^{-3}$ & $1.10 \times 10^{-1}$ & 2.1 & $1.37 \times 10^{-1}$ \\
68d & 0.6 & stress-free & 2.00 & $3.00 \times 10^{6}$ & 2.6 & $1.35 \times 10^{-2}$ & 201 & $1.01 \times 10^{-2}$ & $1.39 \times 10^{-1}$ & 3.6 &  $2.10 \times 10^{-1}$\\
69d & 0.6 & stress-free &2.00 & $4.00 \times 10^{6}$ & 3.5 & 0.04& 284 & $1.42 \times 10^{-2}$ & $1.89 \times 10^{-1}$ & 8.5 & $3.79 \times 10^{-1}$ \\
\midrule
70d & 0.6 & stress-free & 3.00 & $2.50 \times 10^{6}$ & 1.6 & $7.47 \times 10^{-6}$ & 54 & $2.75 \times 10^{-3}$ & $4.35 \times 10^{-2}$ & 0.2 &  $2.09 \times 10^{-2}$\\
71d & 0.6 & stress-free & 3.00 & $2.70 \times 10^{6}$ & 1.8 & $1.72 \times 10^{-5}$ & 71 & $3.59 \times 10^{-3}$ & $5.91 \times 10^{-2}$ & 0.4 &  $3.50 \times 10^{-2}$\\
72d & 0.6 & stress-free & 3.00 & $3.00 \times 10^{6}$ & 2.0 & $4.39 \times 10^{-5}$ & 103 & $5.17 \times 10^{-3}$ & $8.83 \times 10^{-2}$ & 1.1 &  $6.32 \times 10^{-2}$\\
73d & 0.6 & stress-free & 3.00 & $3.20 \times 10^{6}$ & 2.1 & $4.00 \times 10^{-4}$ & 119 & $5.96 \times 10^{-3}$ & $1.00 \times 10^{-1}$ & 1.5 &  $8.05 \times 10^{-2}$\\
74d & 0.6 & stress-free & 3.00 & $3.50 \times 10^{6}$ & 2.3 & $1.08 \times 10^{-3}$ & 140 & $7.05 \times 10^{-3}$ & $1.15 \times 10^{-1}$ & 2.3 &  $1.29 \times 10^{-1}$\\
75d & 0.6 & stress-free &3.00 & $4.00 \times 10^{6}$ & 2.6 & 0.06& 168 & $8.39 \times 10^{-3}$ & $1.38 \times 10^{-1}$ & 12.7 & $5.13 \times 10^{-1}$ \\
76d & 0.6 & stress-free &3.00 & $7.00 \times 10^{6}$ & 4.6 & 0.05& 307 & $1.54 \times 10^{-2}$ & $2.41 \times 10^{-1}$ & 37.5 & $9.86 \times 10^{-1}$ \\
\midrule
\label{tab:results}
\end{longtable}}
}

\end{document}